\documentclass[preprintnumbers,nofootinbib,a4paper]{revtex4-1}

\usepackage{amsmath,latexsym,amssymb,amsfonts}
\usepackage{float} \usepackage{graphicx}
\usepackage{epstopdf}
\usepackage{graphics}
\usepackage{caption}
\usepackage{subfigure}
\usepackage{bm}
\usepackage{color}
\usepackage{hyperref}
\usepackage[dvips]{psfrag}
\usepackage{ulem}
\usepackage{epsfig}

\usepackage[dvipsnames]{xcolor}

\newcommand{\eff}{\text{eff}}
\newcommand{\cro}{\text{cr}}
\newcommand{\DE}{\text{DE}}
\newcommand{\m}{\text{m}}
\newcommand{\tf}{\tilde{f}}
\newcommand{\tF}{\tilde{F}}
\newcommand{\tR}{\tilde{R}}

\newcommand{\mg}{\mathcal{G}}

\newcommand{\tmg}{\tilde{\mathcal{G}}}
\newcommand{\mL}{\mathcal{L}}
\newcommand{\mP}{\mathcal{P}}
\newcommand{\mA}{\mathcal{A}}

\begin{document}

\title{The viable $f(\mg)$ gravity models via reconstruction from the observations}
\author{Seokcheon Lee}
\email[E-mail: ]{skylee2@gmail.com}
\affiliation{Department of Physics, Institute of Basic Science, Sungkyunkwan University, Suwon 16419, Korea}
\author{Gansukh Tumurtushaa}
\email[E-mail: ]{gansuhmgl@ibs.re.kr}
\affiliation{Center for Theoretical Physics of the Universe, Institute for Basic Science (IBS), Daejeon 34051, Korea}
\preprint{CTPU-PTC-20-01}

\begin{abstract}
We reconstruct the viable $f(\mg)$ gravity models from the observations and provide the analytic solutions that well describe our numerical results. In order to avoid unphysical challenges that occur during the numerical reconstruction, we generalize $f(\mg)$ models into $f(\mg_{\mA})$, which is the simple extension of $f(\mg)$ models with the introduction of a constant $\mA$ parameter. We employ several observational data together with the stability condition, which reads $d^2f/d\mg^2>0$ and must be satisfied in the late-time evolution of the universe, to give proper initial conditions for solving the perturbation equation. As a result, we obtain the analytic functions that match the numerical solutions. Furthermore, it might be interesting if one can find the physical origin of those analytic solutions and its cosmological implications. 
\end{abstract}

\maketitle


\section{Introduction}
\label{sec:intro}

Observational evidence~\cite{Riess:1998cb, Perlmutter:1998np} indicates that our present universe has entered into a phase of accelerated expansion. Such an accelerated expansion invokes the existence of a mysterious energy source, dubbed as dark energy~\cite{Perlmutter:1998np}. The existence of dark energy has been independently confirmed by the measurements of Cosmic Microwave Background (CMB) radiation~\cite{Spergel:2003cb, Hinshaw:2012aka, Ade:2013zuv, Ade:2015xua, Aghanim:2018eyx} and the Baryon Acoustic Oscillation (BAO)~\cite{Eisenstein:2005su}. Although the origin of dark energy has not been identified yet, an important quantity describing its property is the equation-of-state parameter (EoS) $\omega_{DE}$, which is very close to $-1$.  

The simplest candidate for dark energy is the cosmological constant $\Lambda$, which represents the vacuum energy density in the $\Lambda$CDM model of the universe. The observed cosmic acceleration of the universe is attributed to the repulsive gravitational force of the $\Lambda$. However, the cosmological constant suffers from the so-called fine-tuning and coincidence problems that respectively refer to the need for explanations to i) why the predicted value of $\Lambda$ if it originates from the vacuum energy in particle physics is much larger than the observed dark energy scale and ii) why the dark matter density is comparable to the vacuum energy density now, given that their time evolution is so different~\cite{Carroll:2000fy, Sahni:1999gb}.  

As an alternative to the cosmological constant, the accelerated expansion of the current universe can also be explained by modifications of the law of gravity at large distances~\cite{DeFelice:2010aj, Linder:2010py, Clifton:2011jh, Mortonson:2013zfa, Koyama:2015vza}. There have been a number of attempts to modify gravity while yielding the late-time acceleration of the universe. One of the simplest is known as the $f(R)$ models of gravity, where $R$ is a Ricci scalar. For a rather minimal modification, one considers that the gravitational Lagrangian may contain some additional terms as $1/R$~\cite{Capozziello:2002rd, Carroll:2003wy}, $\ln R$~\cite{Nojiri:2003ni}, Tr$(1/R)$~\cite{Easson:2005ax}, and inverse powers of Riemann invariant~\cite{Carroll:2004de, Allemandi:2004wn}. Alternatively, it is also possible to take into account the so-called Gauss-Bonnet invariant $\mathcal{G}$ that is a combination of $R$, the Ricci tensor $R_{\mu\nu}$, and the Riemann tensor $R_{\mu\nu\rho\sigma}$ and is expressed as $\mathcal{G}\equiv R^2-4R_{\mu\nu} R^{\mu\nu}+R_{\mu\nu\rho\sigma}R^{\mu\nu\rho\sigma}$. Both the $\mathcal{G}$ and $R$ belong to the so-called Lovelock theories of gravity, an infinite class of curvature invariants, which have an interesting feature that no higher than the second-order derivatives appear in the  equations of motion.  The $f(R, \mathcal{G})$ models of gravity have been previously studied~\cite{Nojiri:2005jg, Cognola:2006eg, Bamba:2009uf, Li:2007jm, DeFelice:2009aj, Cognola:2006sp, Nojiri:2007bt, DeFelice:2008wz, Zhou:2009cy, Uddin:2009wp} to account for not only the late-time cosmic acceleration but also the cosmological viability conditions~\cite{Li:2007jm, Cognola:2006sp, Nojiri:2007bt, DeFelice:2008wz, Zhou:2009cy, Uddin:2009wp} and the solar system constraints~\cite{DeFelice:2009aj}. 

Although the current observations do not have sufficient sensitivity to be able to discriminate dark energy from modified gravity theories, the precise measurement of the large scale structure formation would enable us whether to accept or to rule out the modified gravity scenarios as the origin of the accelerated expansion of the universe. It has therefore been suggested that in order to break the degeneracy between modified gravity models and dark energy, one may utilize the evolution of the linear growth of matter density fluctuations $\delta_{m}(z)=\delta\rho_{m}/\rho_{m}$, where $z$ is the red-shift parameter~\cite{Linder:2004ng, Linder:2007hg, Steigerwald:2014ava, Basilakos:2017rgc}. The dynamical evolution of a small perturbation would be different for different theories of gravity. Thus, it is worth taking the so-called the growth rate factor of matter clustering: $f(a)\equiv d\ln\delta_m(a)/d\ln a \simeq \Omega_m^\gamma(a)$, where the scale factor $a$ is a function of $z$, $\Omega_m(a)$ is the energy fraction of the matter component of the universe, and $\gamma$ is the growth index, into account. The fact that the complexity of both background and perturbation evolution makes it difficult to obtain viable models of modified gravity  that satisfy both the cosmological and local gravity constraints.

The reconstruction technique employed with the observational data in the modified gravity theories is a useful tool on developing viable dark energy models that anticipates the right history of cosmic evolution. Using this scheme, one can compare the corresponding dark energy density with that of the modified gravity one. A number of successful reconstruction methods for $f(R)$ gravity models has been investigated in Refs.~\cite{Carloni:2010ph, He:2012rf, Xu:2014wda, Lee:2017lud}, either by considering the background evolution alone or by adopting the specific models for the reconstruction. An alternative method for reconstructing $f(R)$ models of gravity has been suggested in Ref.~\cite{Lee:2017lud}, in which the equation of state $\omega$ and the growth index $\gamma$ are parameterised as functions of the scale factor and the numerical values provided by the observational data. Thus, based on the formulation introduced in Ref.~\cite{Lee:2017lud}, in this work, we focus on the $f(R, \mathcal{G})$ gravity models. In particular, $R+f(\mathcal{G})$ models of gravity. In our study, we do not specify the forms of $f(\mathcal{G})$. Instead, we aim at reconstructing the $f(\mg)$ models from the observations by using the cosmological parameters and the specific parameterizations of $\omega$ and $\gamma$. In our present study, the equation-of-state and the growth-index parameters take the following forms: $\omega=\omega_0+(1-a)\omega_a$ and $\gamma=\gamma_0+(1-a)\gamma_a$, respectively. Here, the constants $\omega_0$, $\omega_a$, $\gamma_0$, and $\gamma_a$  are supposed to be constrained by observational data~\cite{Lee:2017lud, Chevallier:2000qy, Linder:2002et}.  

This paper is organized as follows. In Sec~\ref{sec:review}, we briefly review the background and the perturbation evolution of the $f(\mg)$ gravity models. In order to prepare our setup for further numerical analyses, we rewrite the necessary equations in terms of the dimensionless quantities in Sec.~\ref{sec:dimensionless}. The Sec.~\ref{sec:connection} serves as the connection between the $f(\mg)$ models and the observations. In this section, we rewrite the background equations for effective dark energy and compare them with the corresponding $f(\mg)$ ones. In Sec.~\ref{sec:reconstruction}, we introduce the generalization of $f(\mg)$ models into the $f(\mg_{\mA})$ models, which ensure the smoothness of the models. We obtain the values of cosmological parameters that well describe the observational data ({\it i.\,e.,} the best-fit values) for three different models in subsection~\ref{subsec:observation}. In subsection~\ref{subsec:observation}, we present our numerical results on the reconstruction of $f(\mg_\mA)$ gravity models and the analytic functions that match with our numerical solutions. We conclude and provide discussions in Sec.~\ref{sec:conclusion}.

\section{Review: Background and Perturbation Evolution}\label{sec:review}

\subsection{Model}\label{subsec:Model}

We start with the action
\begin{equation}\label{eq:action}
S= \frac{c^4}{16 \pi G}\int d^4 x \sqrt{-g}\left[R+ f \left(\mg \right) + \mL_{\m} \right] \,,
\end{equation}
where $R$ is the Ricci scalar, $\mathcal{G}=R^2-4R_{\mu\nu}R^{\mu\nu}+R_{\mu\nu\rho\sigma}R^{\mu\nu\rho\sigma}$ is the Gauss-Bonnet term, $f(\mg)$ is a function of $\mg$, and $\mL_{\m}$ is the Lagrangian of matter fields.  
One can vary the action of Eq.~\eqref{eq:action} with respect to the metric $g_{\mu\nu}$ to obtain the corresponding field equations~\cite{Li:2007jm}
\begin{align}\label{eq:Gmunu}
G_{\mu\nu}-\Sigma_{\mu\nu}= \frac{8\pi G}{c^4} T_{\mu\nu}\,,
\end{align}
where $G_{\mu\nu}=R_{\mu\nu}-g_{\mu\nu}R/2$ is Einstein's tensor and $T_{\mu\nu}$ is the energy-momentum tensor for a perfect fluid. The effective energy-momentum tensor $\Sigma_{\mu\nu}$ is defined as 
\begin{align}
\Sigma_{\mu\nu} &\equiv 4\left[ R_{\mu\rho\sigma \nu}+R_{\mu\rho}g_{\nu\sigma}+R_{\rho\nu}g_{\mu\sigma}-R_{\mu\nu}g_{\rho\sigma}- R_{\rho\sigma}g_{\mu\nu}-\frac{1}{2}R\left( g_{\mu\nu}g_{\rho\sigma}-g_{\mu\sigma}g_{\nu\rho}\right) \right]\nabla^\rho\nabla^\sigma F 
		-\frac{1}{2}\left( \mg F  -f \right)g_{\mu\nu} \label{Sigmamunu} \,. 
\end{align}
where $F = f_{, \mg} =  \partial f/\partial \mg$.
The trace of Eq.~\eqref{eq:Gmunu} is given by
\begin{align}\label{eq:trace}
R + 2f -2\mathcal{G} F -2R \square F +4R_{\mu\nu}\nabla^\mu\nabla^\nu F =- \frac{8\pi G}{c^4} \left(\rho_{\m}-3p_{\m}\right)\, ,
\end{align}
where $\rho_{\m}$ and $p_{\m}$ are the energy density and the pressure of a non-relativistic matter, respectively. Hereafter, we assume that the matter fluid is given under the form of a perfect fluid with a zero pressure and the matter energy density $\rho_{\m}$ satisfies the continuity equation
\begin{align}
\dot{\rho}_{\m}&= -3H \rho_{\m}\,. \label{dotrhom}
\end{align}

\subsection{Background Equations} \label{subsec:background}
In a spatially flat FLRW background with a spacetime metric
\begin{equation}\label{eq:geodesiceq}
ds^2=- c^2 dt^2+a^2(t)d\vec{x}^2\,,
\end{equation}
one can obtain the dynamical equations of motion
\begin{align}
3H^2&= \frac{1}{2} \left( \mg F - f -24 H^3 \dot{F} \right) + \frac{8 \pi G}{c^2} \rho_{\m} \label{Fried1} \,,\\
-2\dot{H}&= 4H\dot{F}\left(2\dot{H}-H^2\right)+4 H^2\ddot{F}+\frac{8\pi G}{c^2} \rho_{\m}\,,\label{Fried2}
\end{align}
where the dot denotes the derivative with respect to  (w.r.t) the cosmic time, $t$.  

It is convenient to replace the time derivatives with the derivatives w.r.t the number of e-folds, $N= \ln a$. Thus, in terms of the $N$, the above background equations can be rewritten as 
\begin{align}
3H^2&= \frac{1}{2} \left( \mg F - f -24 H^4 F' \right) + \frac{8 \pi G}{c^2} \rho_{\m} \label{pFried1} \,,\\
-2 \frac{H'}{H} &= 4H^2 F' \left(2 \frac{H'}{H} - 1 \right)+4 H^2 \left( F'' + \frac{H'}{H} F' \right) + \frac{8\pi G}{c^2} \frac{1}{H^2} \rho_{\m}\,,\label{pFried2} 
\end{align}
where the prime denotes the derivatives w.r.t the $N$. In this flat background, the Ricci scalar and the Gauss-Bonnet term are given by 
\begin{align}\label{eq:Ricci}
R&=6(2H^2+\dot{H}) = 6 H^2 \left( 2 + \frac{H'}{H} \right) \,,\\
\mg&=24H^2(H^2+\dot{H}) = 24 H^4 \left( 1 + \frac{H'}{H} \right) \,. \label{eq:GBterm}
\end{align}

\subsection{Perturbations Equation}\label{subsec:perturbation}

For the sub-horizon modes ($c^2k^2\gg a^2H^2$), the evolution equation for the matter perturbation is given by \cite{DeFelice:2010hb}
\begin{equation}
\ddot{\delta}_m+2H\dot{\delta}_m-4\pi G \rho_m \left[ \frac{A_1+A_2\left(\frac{{\color{black} c} k}{a H}\right)^2}{B_1+B_2\left(\frac{{\color{black} c} k}{a H}\right)^2}\right]\delta_m=0\,, \label{Pert1}
\end{equation}
where 
\begin{align}
A_1 &= 1 + 4 \ddot{F} \label{A1} \,,\\
A_2 &
= 64 H^2 \frac{\dot{F}}{\dot{\mg}} \left(\dot{H}+H^2\right)^2 \label{A2} \,,\\
B_1 &= \left( 1 + 4 H \dot{F} \right)^2 \label{B1} \,,\\
B_2 
	&= 16 H^4 \frac{\dot{F}}{\dot{\mg}} \left[\left(4+16H\dot{F} \right)\left( \dot{H}+H^2\right)-H^2(1+4\ddot{F} )\right] \label{B2} \, .
\end{align}
Eq.~(\ref{Pert1}) 
can also be rewritten in terms of $N$ as follows:
\begin{align}\label{Pert2}
\delta''_m+\left(2+\frac{H'}{H}\right)\delta'_m = \frac{3}{2}\Omega_{m}\left[ \frac{A_1+A_2\left(\frac{ck}{a H}\right)^2}{B_1+B_2\left(\frac{ck}{a H}\right)^2}\right]\delta_m\,,
\end{align}
where $\Omega_{m} \equiv(8 \pi G/c^2) \rho_{\m}/(3H^2)$ and 
\begin{align}
A_1 &= 1 + 4 H^2 \left( F^{\prime\prime} + \frac{H'}{H} F' \right)  \label{A1p} \,, \\
A_2 &= 64 H^{6} \frac{F'}{\mg'} \left( \frac{H'}{H} + 1 \right)^2 \label{A2p} \,, \\
B_1 &= \left(1+4H^2 F' \right)^2 \label{B1p} \,, \\
B_2 &= 16 H^6 \frac{F'}{\mg'} \left[\left(4+16 H^2 F' \right)\left( \frac{H'}{H} + 1 \right) - \left(1+4 \frac{H'}{H} H^2 F' + 4 H^2 F^{\prime\prime} \right)\right] \label{B2p}  \,.
\end{align}
For the wavenumber, $k$ which has units of $[h/\text{Mpc}]$, dependent term in the square parenthesis, we use the following expression  
\begin{align}
\frac{ck}{aH} &= \frac{ck}{a_0 H_0} \frac{a_0 H_0}{a H} = 3000 k \frac{H_0}{H} e^{-N} \label{ckoaH} \, , 
\end{align}
where the current value of the scale factor of the universe is normalized to unity ({\it i.e.,} $a_0 = 1$). 
It is worth noting here that the term inside the square parenthesis in Eqs.~(\ref{Pert1}) and (\ref{Pert2}) reflects deviation from GR and is a function of both the wavenumber $k$ and the scale factor $a$ ({\it i.e.,} in the limit $f(\mathcal{G})\rightarrow\text{const.}$ (or $0$), $A_1=1$, $A_2=0$, $B_1=1$, and $B_2=0$, the GR is recovered). For the models of $f(\mg)$ whose deviation from the GR is small during radiation- and matter-dominated eras, $f_{,\mg\mg}\equiv d^2f(\mg)/d\mg^2$ is closer to zero. However, in order for not to violate the stability of perturbations, we require the condition that reads $f_{,\mg\mg}> 0$~\cite{DeFelice:2008wz}.

\section{Equations in terms dimensionless quantities}
\label{sec:dimensionless}

One can notice from Eq.~(\ref{eq:action}) that the dimension of $f(\mg)$ should be the same as that of $R$. Thus, if one normalizes $f(\mg)$ by $H_0^2$ then it becomes a dimensionless quantity. Hereafter, in our upcoming numerical calculations, we will treat $f/H_0^2 \equiv \tf $ as a number. Similarly, from Eqs.\eqref{eq:Ricci}-\eqref{eq:GBterm}, we define the dimensionless quantities for the Ricci scalar and the Gauss-Bonnet term as follows
\begin{align}\label{tRicci}
\tR &\equiv \frac{R}{H_0^2} = 6 \left(\frac{H}{H_0} \right)^2 \left( 2 + \frac{H'}{H} \right) \,,\\
\tmg &\equiv \frac{\mg}{H_0^4} = 24 \left(\frac{H}{H_0} \right)^4 \left( 1 + \frac{H'}{H} \right) \,. \label{tGBterm}
\end{align}
Therefore, in order to perform the numerical reconstruction of models, we need to rewrite both the background and the perturbation equations in terms of dimensionless quantities. First, the background evolution equations Eq~\eqref{pFried1} and \eqref{pFried2} read
\begin{align}
\frac{H^2}{H_0^2} &= -4 \frac{H^4}{H_0^4} \frac{1}{\tmg'} \left[ \tf'' - \left( \frac{\tmg''}{\tmg'} + \frac{1}{24} \frac{H_0^4}{H^4} \tmg \right) \tf' + \frac{1}{24} \frac{H_0^4}{H^4} \tmg' \tf \right]+ \Omega_{\m0} e^{-3 N} \label{pFried1dim} \,, \\
\frac{H'}{H}
&= -2 \frac{H^2}{H_0^2} \frac{1}{\tmg'} \left[ \tf''' + \left( 3 \frac{H'}{H} - 2 \frac{\tmg''}{\tmg'} - 1 \right) \tf'' - \left( \frac{\tmg'''}{\tmg'} + \left( 3 \frac{H'}{H} - 2 \frac{\tmg''}{\tmg'} - 1 \right) \frac{\tmg''}{\tmg'}  \right) \tf' \right] -\frac{3}{2} \frac{H_0^2}{H^2} \Omega_{m0}  e^{-3 N} \label{pFried2dim} \,.
\end{align}
Here, the tilde indicates corresponding dimensionless quantities, such as $\tilde{f} \equiv f/H_0^2$, $\tmg \equiv \mg/H_0^4$, $\tilde{f}^{(n)} \equiv d^{n} \tilde{f}/ dN^{n}$, and $\tmg^{(n)} \equiv d^{n} \tmg/ dN^{n}$ where we use following chain rule: 
\begin{align}\label{tildeF} 
\tilde{F} &= \frac{1}{\tmg'} \tilde{f}' \, , \quad  \tilde{F}^{\prime}  = \frac{1}{\tmg'} \left( \tilde{f}^{\prime\prime} - \frac{\tmg^{\prime\prime}}{\tmg^{\prime}} \tilde{f}^{\prime} \right) \,, \quad \tilde{F}^{\prime\prime} = \frac{1}{\tmg'} \left\{ \tilde{f}^{\prime\prime\prime} - 2 \frac{\tmg^{\prime\prime}}{\tmg^{\prime}} \tilde{f}^{\prime\prime}  + \left[ 2 \left( \frac{\tmg^{\prime\prime}}{\tmg^{\prime}} \right)^{2} -  \frac{\tmg^{\prime\prime\prime}}{\tmg^{\prime}} \right] \tilde{f}^{\prime} \right\}  \, . 
\end{align} 

Next, one can also rewrite the equation of the matter perturbation in Eq.~\eqref{Pert2} by using the dimensionless quantities
\begin{align}
\frac{\delta''_{\m}}{\delta_{\m}} &+\left(2+\frac{H'}{H}\right) \frac{\delta'_{\m}}{\delta_{\m}} = \frac{3}{2}\Omega_{m}\left[ \frac{A_1+A_2\left(\frac{ck}{a H}\right)^2}{B_1+B_2\left(\frac{ck}{a H}\right)^2}\right] \equiv \mP \label{Pert2dim} \,,
\end{align}
where 
\begin{align}
A_1 
	&\equiv 1 +  A_{1}^{(3)} \tf''' + A_{1}^{(2)} \tf'' + A_{1}^{(1)} \tf' \,, \\
A_2 
	&\equiv A_{2}^{(2)} \tf'' + A_{2}^{(1)} \tf' \, , \\
B_1 
	&\equiv 1 + B_{1}^{(2)} \tf'' + B_{1}^{(1)} \tf' + B_{1}^{(22)} \tf^{\prime\prime 2} + B_{1}^{(21)} \tf'' \tf' + B_{1}^{(12)} \tf^{\prime 2} \,, \label{B1dim} \\ 
B_2 
&\equiv B_{2}^{(2)} \tf'' +  B_{2}^{(1)} \tf' + B_{2}^{(32)} \tf''' \tf'' + B_{2}^{(31)} \tf''' \tf' +  B_{2}^{(22)} \tf^{\prime\prime 2} + B_{2}^{(21)} \tf'' \tf' + B_{2}^{(12)} \tf^{\prime 2} \,, \label{B2dim}
\end{align} 
with the coefficients
\begin{align}
A_{1}^{(3)} & =   \frac{4}{\tmg'}\frac{H^2}{H_0^2}\,,\,\,\, A_{1}^{(2)} = \frac{4}{\tmg'}\frac{H^2}{H_0^2} \left( \frac{H'}{H} - 2 \frac{\tmg''}{\tmg'} \right)\,, \,\,\, A_{1}^{(1)} = -\frac{4}{\tmg'}\frac{H^2}{H_0^2} \left[ \frac{H'}{H} \frac{\tmg''}{\tmg'} - 2 \left( \frac{\tmg''}{\tmg'} \right)^2 + \frac{\tmg'''}{\tmg'} \right]\,,\\
A_{2}^{(2)} & = \frac{64}{\tmg^{\prime 2}}  \left( \frac{H^{2}}{H_0^2} \right)^{3}  \left( \frac{H'}{H} + 1 \right)^2 \,, \quad A_{2}^{(1)} = - \frac{64}{\tmg^{\prime 2}}  \left( \frac{H^{2}}{H_0^2} \right)^{3}  \left( \frac{H'}{H} + 1 \right)^2 \left(\frac{\tmg''}{\tmg'} \right)\,,\\
B_1^{(2)} &=  \frac{8}{\tmg'} \frac{H^2}{H_0^2}\,, \,\,\, B_1^{(1)} =  -\frac{8}{\tmg'} \frac{H^2}{H_0^2} \frac{\tmg''}{\tmg'}\,,\,\,\, B_1^{(22)} = \left(\frac{4}{\tmg'} \frac{H^2}{H_0^2}\right)^2\,,\,\,\, B_1^{(21)} = -2\left(\frac{4}{\tmg'} \frac{H^2}{H_0^2}\right)^2\frac{\tmg''}{\tmg'} \,,\,\,\, B_1^{(12)} = \left(\frac{4}{\tmg'} \frac{H^2}{H_0^2}\frac{\tmg''}{\tmg'}\right)^2\,,\\
B_{2}^{(2)} &= \left(\frac{4}{\tmg^{\prime}}  \frac{H^3}{H_0^3}\right)^2 \left( 3 + 4 \frac{H'}{H} \right) \,,\,\,\, B_{2}^{(1)} = - \left(\frac{4}{\tmg^{\prime}}  \frac{H^3}{H_0^3}\right)^2 \left( 3 + 4 \frac{H'}{H} \right) \frac{\tmg^{\prime\prime} }{\tmg^{\prime}}\,,\,\,\, B_{2}^{(32)} = - \frac{64}{\tmg^{\prime 3}} \frac{H^8}{H_0^8} \,,\,\,\, B_{2}^{(31)} = \frac{64}{\tmg^{\prime 3}} \frac{H^8}{H_0^8}\frac{\tmg''}{\tmg'}\,,\nonumber \\
B_{2}^{(22)} &= \frac{64}{\tmg^{\prime 3}} \frac{H^8}{H_0^8} \left(4 + 3 \frac{H'}{H} + 2 \frac{\tmg''}{\tmg'} \right) \,,\,\,\, B_{2}^{(21)} = - \frac{64}{\tmg^{\prime 3}} \frac{H^8}{H_0^8} \left[2 \frac{\tmg''}{\tmg'} \left(4 + 3 \frac{H'}{H} + 2 \frac{\tmg''}{\tmg'}  \right) - \frac{\tmg'''}{\tmg'} \right]  \,, \nonumber \\
B_{2}^{(12)} &=  \frac{64}{\tmg^{\prime 3}} \frac{H^8}{H_0^8} \left[ \left(\frac{\tmg''}{\tmg'}\right)^2 \left( 4 + 3 \frac{H'}{H} + 2 \frac{\tmg''}{\tmg'}\right) - \frac{\tmg'''}{\tmg'} \right] \,.
\end{align}
As is seen in Eqs.~\eqref{B1dim} and \eqref{B2dim}, both $B_1$ and $B_2$ include the multiplication of derivatives of $\tf$. Thus, it is safe for us to ignore those terms in our numerical analysis as long as the assumption of slowly varying $\tf$ is satisfied.

In the following section, we show that $(H/H_0)^2$, $H'/H$, $\delta_{\m}'/\delta_{\m}$, and $\delta_{\m}''/\delta_{\m}$ can be obtained from cosmological observations. In other words, they can be expressed in terms of observable quantities. Thus, one can obtain the time evolution of $f(\tmg)$ function for the given values of cosmological parameters obtained from observations. For this purpose, we combine Eqs.\eqref{pFried1dim} and \eqref{pFried2dim} to obtain

\begin{align}
& \tf''' + \left( 3 \frac{H'}{H} -2 \frac{\tmg''}{\tmg'} + 2 \right) \tf'' - \left[ \frac{\tmg'''}{\tmg'} + \frac{\tmg'}{8} \frac{H_0^4}{H^4}  + \left( 3 \frac{H'}{H} -2 \frac{\tmg''}{\tmg'} + 2 \right) \frac{\tmg''}{\tmg'} \right] \tf' + \frac{\tmg'}{8} \frac{H_0^4}{H^4}  \tf 
= -\frac{\tmg'}{2} \frac{H_0^2}{H^2} \left( \frac{3}{2} + \frac{H'}{H} \right) \label{MEdimf} \,.
\end{align} 
As we can see, Eq.\eqref{MEdimf} is a third-order-linear-inhomogeneous differential equation for $f(N)$. Thus, as long as the initial conditions for $\tilde{f}(N)$, $\tilde{f}^{\prime}(N)$, and $\tilde{f}^{\prime\prime}(N)$ is given, one can solve the above equation. In general, we need four constraint equations to specify initial conditions completely.  Thus, to obtain these initial conditions, we use both the background and the perturbation equations.   
By employing the background evolution equations given in Eqs.~\eqref{pFried1dim} and \eqref{pFried2dim} together with the perturbation equations given in Eq.~\eqref{Pert2dim} at the present time, we find equations for the initial conditions as 
%
\begin{align}
&\tf_0^{\prime\prime} - \left( \frac{\tmg_0^{\prime\prime}}{\tmg_0^{\prime}} + \frac{\tmg_0}{24} \right) \tf_{0}^{\prime} + \frac{\tmg_{0}'}{24} \tf_{0} = -\frac{\tmg_0'}{4} \left( 1 - \Omega_{\m 0} \right) \label{ini1-1} \, , \\
&\tf_{0}^{\prime\prime\prime} - \left( 2 \frac{\tmg_{0}^{\prime\prime}}{\tmg_{0}^{\prime}} + \frac{H_0^{\prime}}{H_0} \right) \tf_{0}^{\prime\prime} - \left[ \frac{\tmg_{0}^{\prime\prime\prime}}{\tmg_{0}^{\prime}} - \frac{H_0^{\prime}}{H_0}  \frac{\tmg_{0}^{\prime\prime}}{\tmg_{0}^{\prime}}  - 2 \left( \frac{\tmg_{0}^{\prime\prime}}{\tmg_{0}^{\prime}} \right)^{2} \right] \tf_{0}^{\prime} + \left( 1 - 2 \frac{H_0^{\prime}}{H_0}  \right) \tf_{0} &= -\frac{3}{4} \left( 1 - \Omega_{\m 0} \right) \left( 1 + \omega_{\DE 0} \right) \tmg_{0}^{\prime} \label{ini2-1} \, , 
\end{align} 

\begin{align}
\frac{2}{3} \frac{{\cal P}_{0} }{\Omega_{\m0}} - 1 = & A_{10}^{(3)} \tf_{0}'''+\left[ A_{10}^{(2)} - B_{10}^{(2)} + \left( A_{20}^{(2)} - B_{20}^{(2)} \right) \left( \frac{ck}{a_0H_0}\right)^{2} \right] \tf_{0}'' +\left[ A_{10}^{(1)} - B_{10}^{(1)} + \left( A_{20}^{(1)} - B_{20}^{(1)} \right) \left( \frac{ck}{a_0H_0}\right)^{2} \right] \tf_{0}'  
\label{ini3}\,,
\end{align}
where the subscript ``0" denotes the present time value of each quantity.

We show in the next section that the necessary functions can be obtained from cosmological observation hence they are given in terms of cosmological parameters including $\omega_0, \omega_a$, and $\Omega_{m0}$. Thus, $\tilde{f}_0$, $\tilde{f}_0^{\prime}$, and $\tilde{f}_0^{\prime\prime}$ with be given with reasonable initial values.

\section{Connection to observation}
\label{sec:connection}

In principle, one can rewrite the background equations given in Eqs.\eqref{pFried1} and \eqref{pFried2} by using the effective dark energy (EDE) under the assumption that contributions of $f(\mg)$ are those of the EDE
\begin{align}
3H^2&= \frac{1}{2}\left(\mathcal{G} F-f-24H^4 F'\right)+\frac{8\pi G}{c^2}\rho_{\m}  \equiv \frac{8\pi G}{c^2} \left(\rho_{\eff}+\rho_{\m} \right) \, \label{pFried1eff} \\
-2 H H'&= 4H^{4} F'\left(3 \frac{H'}{H} - 1 \right)+4 H^4F'' + \frac{8\pi G}{c^2} \rho_{\m} \equiv \frac{8\pi G}{c^2} \left( \rho_{\eff}+p_{\eff}+\rho_{\m} \right) \,\label{pFried2eff} 
\end{align}  
where $\rho_{\cro 0}$ denotes the critical energy density at present and the energy density, the pressure, and the equation of state of the EDE are given by

\begin{align}\label{rhoeff}
\rho_{\eff} 
&= -\frac{4 \rho_{\cro 0}}{\tmg'}  \frac{H^4}{H_0^4}\left[ \tf'' - \left( \frac{\tmg''}{\tmg'} + \frac{H_0^4}{H^4} \frac{\tmg}{24} \right) \tf' + \frac{H_0^4}{H^4} \frac{\tmg'}{24} \tf \right] \,, \\
p_{\eff} 
&= \frac{4 \rho_{\cro 0}}{3\tmg'}  \frac{H^4}{H_0^4}\left[ \tf''' - \left( 2 \frac{\tmg''}{\tmg'} - 3 \frac{H'}{H} -2 \right) \tf'' - \left[ \frac{\tmg'''}{\tmg'} - \left( 2 \frac{\tmg''}{\tmg'} - 3 \frac{H'}{H} -2 \right) \frac{\tmg''}{\tmg'} + \frac{\tmg}{8}\frac{H_0^4}{H^4} \right] \tf' + \frac{\tmg'}{8} \frac{H_0^4}{H^4} \tf \right] \,,\label{peff} \\
\omega_{\eff}&\equiv \frac{p_{\eff}}{\rho_{\eff}} 
= -1 - \frac{\tf''' - \left( 2 \frac{\tmg''}{\tmg'} - 3 \frac{H'}{H} + 1 \right) \tf'' - \left[ \frac{\tmg'''}{\tmg'} - \left( 2 \frac{\tmg''}{\tmg'} - 3 \frac{H'}{H} + 1 \right) \frac{\tmg''}{\tmg'} \right] \tf' }{ 3 \tf'' - \left( 3 \frac{\tmg''}{\tmg'} + \frac{H_0^4}{H^4} \frac{\tmg}{8} \right) \tf' + \frac{H_0^4}{H^4} \frac{\tmg'}{8} \tf  } \,.\label{omegaeff}
\end{align}
Following the method discussed in Ref.~\cite{Lee:2017lud}, we aim at reconstructing the general $R+f(\mathcal{G})$ models from observations. It is therefore efficient to adopt the parametrizations of cosmological parameters in order to probe various theoretical models. For this purpose, we adopt the so-called Chevalllier-Polarski-Linder (CPL) parameterization of the EDE equation of state: $\omega_{\DE}=\omega_0+\omega_a(1-a)$~\cite{Chevallier:2000qy}. Thus, background evolution equations of motion Eqs.~\eqref{Fried1} and~\eqref{Fried2} can be rewritten as 
\begin{align}
3H^2&= \frac{8\pi G}{c^2} \left( \rho_{\DE}+\rho_{\m} \right) \equiv \frac{8\pi G}{c^2} \rho_{\cro}\,,\label{Fried1DE}\\
-2HH'&= \frac{8\pi G}{c^2} \left( \rho_{\DE}+p_{\DE}+\rho_{\m} \right)\,,\label{Fried2DE}
\end{align}
where $\rho_{\cro}$ is the critical energy density of the universe and 
\begin{align}
\rho_{\DE}&= \rho_{\DE 0} e^{-3(1+\omega_0+\omega_a)N - 3\omega_a (1- e^N)} \label{rhoDE} \,,\\
p_{\DE}&= \omega_{\DE}\rho_{\DE} \label{pDE} \,.
\end{align} 
The values of $\omega_{0}$, $\omega_a$, and $\Omega_{\m 0}$ in Eqs.~\eqref{Fried1DE} and ~\eqref{Fried2DE} can be obtained from cosmological observations and the best-fit values of these parameters are supposed to be used for reconstructing theoretical models of $R+f(\mg)$ by replacing $\rho_{\eff}$ and $p_{\eff}$ with $\rho_{\DE}$ and $p_{\DE}$.  Thus, the obtained value of $\omega_{\eff}$ can be different from $\omega_{\DE}$. The reconstructed models can be accepted as long as this difference in $\omega$ values within the measurement error.

The matter and dark energy components of the universe can also be expressed in terms of these measured quantities as follows:
\begin{align}
\Upsilon[\Omega_{\m 0}, \omega_0,\omega_a, N] &\equiv \frac{\Omega_{\DE}[\Omega_{\m 0}, \omega_0,\omega_a, N] }{\Omega_{\m}[\Omega_{\m 0}, \omega_0,\omega_a, N] } =\frac{1-\Omega_{\m 0}}{\Omega_{\m 0}}e^{-3(\omega_0+\omega_a)N-3\omega_a(1-e^N)} \label{gDE} \,,\\
\Omega_{\m} [\Omega_{\m0}, \omega_0,\omega_a, N]&= \left( 1+ \Upsilon[\Omega_{\m0}, \omega_0,\omega_a, N] \right)^{-1} \label{OmDE} \,,\\
\Omega_{\DE}[\Omega_{\m0}, \omega_0,\omega_a, N]&\equiv 1-\Omega_{\m} [\Omega_{\m0}, \omega_0,\omega_a, N]=\frac{\Upsilon[\Omega_{\m0}, \omega_0,\omega_a, N]}{1+\Upsilon[\Omega_{\m0}, \omega_0,\omega_a, N]}\,.\label{ODEDE}
\end{align}
The Friedmann equations written in Eqs.~\eqref{Fried1DE} and~\eqref{Fried2DE} therefore become
\begin{align}
\frac{H^2}{H^2_0}&=\frac{\rho_{\m}}{\rho_{\cro0}}\left(1+\frac{\rho_{\DE}}{\rho_{\m}}\right)=\Omega_{\m0}\left( 1+\Upsilon[\Omega_{\m0}, \omega_0,\omega_a, N]\right)e^{-3N} \label{Fried1DE-2}\,,\\
\frac{H'}{H}&= -\frac{3}{2}\left(1+\omega_{\DE}\Omega_{\DE}\right)\equiv -\frac{3}{2}\left(1+Q[\Omega_{\m0}, \omega_0,\omega_a, N]\right) \label{Fried2DE-2} \,,
\end{align}
where 
\begin{align} 
Q[\Omega_{\m0}, \omega_0,\omega_a, N]=\left(\frac{\Upsilon[\Omega_{\m0}, \omega_0,\omega_a, N]}{1+\Upsilon[\Omega_{\m0}, \omega_0,\omega_a, N]}\right)\omega_{\DE}[\Omega_{\m0}, \omega_0,\omega_a, N] \,.
\end{align} 
In addition, Eqs.~\eqref{tRicci} and \eqref{tGBterm} are written as
\begin{align}
\tR &=  3\Omega_{\m0}(1+\Upsilon)(1-3Q)e^{-3N} \label{tRicciDE} \,,\\
\tmg&=  -12\Omega_{\m0}^2(1+\Upsilon)^2(1+3Q) e^{-6N} \label{tGBtermDE} \,.
\end{align}
The dark energy density given in Eq.~\eqref{ODEDE} and its equation of state $\omega_{DE}$ can be compared with those derived from Eqs.~\eqref{rhoeff} and \eqref{omegaeff} for the $R+f(\mathcal{G})$ models,

\begin{align}
\Omega_{\DE} \simeq \Omega_{\eff}: \quad & 
\frac{\Upsilon[\Omega_{\m0}, \omega_0,\omega_a, N]}{1+\Upsilon[\Omega_{\m0}, \omega_0,\omega_a, N]} 
	\simeq\frac{-4  \frac{H^4}{H_0^4} \frac{1}{\tmg'} \left[ \tf'' - \left( \frac{\tmg''}{\tmg'} + \frac{H_0^4}{H^4} \frac{\tmg}{24} \right) \tf' + \frac{H_0^4}{H^4} \frac{\tmg'}{24} \tf \right] }{ -4 \frac{H^4}{H_0^4} \frac{1}{\tmg'} \left[ \tf'' - \left( \frac{\tmg''}{\tmg'} + \frac{H_0^4}{H^4} \frac{\tmg}{24} \right) \tf' + \frac{H_0^4}{H^4} \frac{\tmg'}{24} \tf \right] + \Omega_{\m0} e^{-3N}}\,, \label{OmegaDEeff} \\
\omega_{\DE} \simeq \omega_{\eff}: \quad  & 
\omega_{\DE}[\omega_0,\omega_a, N] \simeq  -1 - \frac{\tf''' - \left( 2 \frac{\tmg''}{\tmg'} - 3 \frac{H'}{H} + 1 \right) \tf'' - \left( \frac{\tmg'''}{\tmg'} - \left( 2 \frac{\tmg''}{\tmg'} - 3 \frac{H'}{H} + 1 \right) \frac{\tmg''}{\tmg'} \right) \tf' }{ 3 \tf'' - \left( 3 \frac{\tmg''}{\tmg'} + \frac{H_0^4}{H^4} \frac{\tmg}{8} \right) \tf' + \frac{H_0^4}{H^4} \frac{\tmg'}{8} \tf  } \,. \label{omegaDEeff}
\end{align}
The same scheme can be applied to the equation for the matter perturbation given in Eq.~\eqref{Pert2dim}. The growth rate of the matter perturbation is well parametrized as 
\begin{align}
\Omega_{\m}^{\gamma} \equiv  \frac{ d \ln \delta_{\m}}{ d \ln a} = \frac{\delta_{\m}'}{\delta_{\m}}\,, \label{Omgamma} 
\end{align}
where $\gamma$ is the growth-rate index and we use the following parameterization: $\gamma \equiv \gamma_0 + \gamma_{a} \left( 1 - e^{N} \right)$. Here, the values of $\gamma_0$ and $\gamma_a$ are to be provided by observational constraints. By using Eq.~\eqref{Omgamma}, one can rewrite the left-hand side of Eq.~\eqref{Pert2dim} as a function of cosmological parameters
\begin{align}
\mathcal{P}[\Omega_{\m0}, \omega_0,\omega_a, \gamma_0, \gamma_a, N] 
&= (1+\Upsilon)^{-\gamma}\left[ (1+\Upsilon)^{-\gamma} -\gamma' \ln(1+\Upsilon)+ 3 \gamma Q  + \frac{1}{2} (1-3Q)\right] \label{P} \,.
\end{align}
By using  Eq.~\eqref{P}, we write Eq.~\eqref{Pert2dim} as 
\begin{align}
\frac{2}{3} \frac{\mathcal{P}[\Omega_{\m0}, \omega_0,\omega_a, \gamma_0, \gamma_a, N]}{\Omega_{\m}[\Omega_{\m0}, \omega_0,\omega_a, N]} 
	\simeq &\,\, 1 + A_{1}^{(3)} \tf''' + \left[ A_{1}^{(2)} - B_{1}^{(2)} + \left( A_{2}^{(2)} - B_{2}^{(2)}\right) \left( \frac{ck}{H_0} \right)^2 \left( \frac{H_0^2}{H^2} \right) e^{-2N} \right] \tf'' \nonumber \, \\
	&\,\, +  \left[ A_{1}^{(1)} - B_{1}^{(1)} + \left( A_{2}^{(1)} - B_{2}^{(1)}\right) \left( \frac{ck}{H_0} \right)^2 \left( \frac{H_0^2}{H^2} \right) e^{-2N} \right] \tf' \,.
\end{align}

Now we are ready to obtain all the necessary initial conditions for solving Eq.~\eqref{MEdimf} from Eqs.~\eqref{ini1-1}, \eqref{ini2-1}, and \eqref{ini3}
\begin{align}
1 &= -4 \frac{1}{\tmg_0'} \left[ \tf_0'' - \left( \frac{\tmg_0''}{\tmg_0'} + \frac{1}{24} \tmg_0 \right) \tf_0' + \frac{1}{24} \tmg_0' \tf_0 \right]+ \Omega_{\m0}  \label{pFried1dimini} \,, \\
\frac{H_0'}{H_0} &= -2 \frac{1}{\tmg_0'} \left[ \tf_0''' + \left( 3 \frac{H_0'}{H_0} - 2 \frac{\tmg_0''}{\tmg_0'} - 1 \right) \tf_0'' - \left( \frac{\tmg_0'''}{\tmg_0'} + \left( 3 \frac{H_0'}{H_0} - 2 \frac{\tmg_0''}{\tmg_0'} - 1 \right) \frac{\tmg_0''}{\tmg_0'}  \right) \tf_0' \right] -\frac{3}{2} \Omega_{\m0}  \label{pFried2dimini} \, \\
	&= -\frac{3}{2} \left( 1 + Q_0 \right) =  -\frac{3}{2} \left( 1 + \omega_{\DE0} \Omega_{\DE 0} \right) \nonumber \,, \\
\frac{2}{3} \frac{{\cal P}_{0}}{\Omega_{\m0}} - 1 &=  A_{10}^{(3)} \tf_0''' + \left[ A_{10}^{(2)} - B_{10}^{(2)} + \left( A_{20}^{(2)} - B_{20}^{(2)}\right) \left( \frac{ck}{H_0} \right)^2 \right] \tf_0'' +  \left[ A_{10}^{(1)} - B_{10}^{(1)} + \left( A_{20}^{(1)} - B_{20}^{(1)}\right) \left( \frac{ck}{H_0} \right)^2 \right] \tf_0' \nonumber \, \\
	&= \frac{2}{3} \Omega_{\m0}^{\gamma_0 - 1} \left[ \Omega_{\m0}^{\gamma_0} - \gamma_{a} \ln \Omega_{\m0} + 3 \gamma_0 Q_{0} + \frac{1}{2} \left( 1 - 3 Q_0 \right)\right] - 1 \label{PoOmini}  \,, 
\end{align}
where the following necessary functions
\begin{eqnarray}
\frac{H''}{H}&=&-\frac{3}{2}\left[Q'-\frac{3}{2}(1+Q)^2 \right]\,,\\
\frac{R'}{H_0^2}&=&-9 \Omega_{m0}(1+\Upsilon)\left[Q'+(1+Q)(1-3Q) \right]e^{-3N}\,,\\
\frac{R''}{H_0^2}&=&-9\Omega_{m0}(1+\Upsilon)\left[Q''-(9Q+5)Q'-3(1+Q)^2(1-3Q) \right] e^{-3N}\,,\\
\frac{\mathcal{G}'}{H_0^4}&=&-36\Omega_{m0}^2(1+\Upsilon)^2 \left[Q'-2(1+Q)(1+3Q)\right] e^{-6N}\,,\\
\frac{\mathcal{G}''}{H_0^4}&=&-36 \Omega_{m0}^2(1+\Upsilon)^2\left[Q''-2(9Q+7)Q'+12(1+Q)^2(1+3Q) \right]e^{-6N}\,.
\end{eqnarray}
can be evaluation at $N=0$. 

\section{Observational constraints and reconstructed $f(\mg)$ models}\label{sec:reconstruction}

Before we probe details of reconstruction of $f(\mg)$ models, it is worth emphasizing the differences between $f(R)$ models and $f(\mg)$ models. As one can see in Eqs.~\eqref{tRicci} and \eqref{tGBterm}, both depend on $H'/H$ term and this can vary from $-3/2$ to $0$ depending on an equation of state, see Eq.~\eqref{Fried2DE-2}. The Ricci scalar, $R$ is proportional to $\left( 2 + H'/H \right)$ and thus $R$ and its derivatives do not change signs during their evolutions. However, $\mg$ is proportional to $\left( 1 + H'/H \right)$ and both $\mg$ and its first-order derivative with respect to $N$ change their signs during their cosmological evolutions. These change in sign of the first-order derivative of $\mg$ causes the divergence in the numerical work and also make trouble in the interpretation of physical quantities related to $\mg'$. Thus, it is troublesome to use a general form of $f(\mg)$ as one does in $f(R)$ gravity models. In order to avoid this singularity problem in $f(\mg)$ models, one can adopt the simple extension of it as
\begin{align}
\mg_{\mA} &\equiv \mg + \mA H^4 = 24 H^4 \left( \mA + 1 + \frac{H'}{H} \right)  \label{mganal} \,,   \\
\tmg_{\mA} &\equiv \tmg + \mA \left( \frac{H}{H_0} \right)^4 = 24 \left( \frac{H}{H_0} \right)^4 \left( \mA + 1 + \frac{H'}{H} \right) \label{tmganal} \,,
\end{align}  
where $\mA>1/2$ is the dimensionless constant. In general, $f(R)$ models do not have this kind constraint but so do $f(\mg)$. Of course, this extension is not unique and there can be various extensions of $f(\mg)$. However, we limit ourselves to this simplest extension model, $f(\mg_{\mA})$ in the rest of this manuscript. \\

\subsection{Current observational constraints on cosmological parameters}\label{subsec:observation}

In this subsection, we present the observational constraints on the cosmological parameters: $\Omega_{m0}$, $\omega_0$, $\omega_a$, $\gamma_0$, and $\gamma_a$, discussed in the preceding sections. In our analysis, we use observational data including the CMB\cite{Ade:2015xua, Wang:2015tua}, Supernovae type Ia (SnIa)~\cite{Scolnic:2017caz}, BAO~\cite{Beutler:2011hx, Xu:2012hg, Anderson:2013zyy, Ross:2014qpa, Gil-Marin:2015nqa}, Hubble expansion $H(z)$~\cite{Moresco:2016mzx, Guo:2015gpa}, and the growth-rate data~\cite{Sagredo:2018ahx, Amendola:2012wc}.  The total likelihood function $\mL_{tot}$ can, therefore, be given as the product of the separate  likelihoods of each data as follows: $\mL_{tot} = \mL_{CMB} \times \mL_{SnIa} \times \mL_{BAO} \times \mL_{H(z)} \times \mL_{growth}$, 
which is also related to the total $\chi^2$ via $\chi_{tot}^2=-\log \mL_{tot}$ or $\chi_{tot}^2 = \chi_{CMB}^2+\chi_{SnIa}^2+\chi_{BAO}^2+\chi_{H(z)}^2+\chi_{growth}^2$. By employing the aforementioned cosmological data together with the statistical methods of minimizing the $\chi_{tot}^2$, we can obtain the best-fit values of the cosmological parameters $\{\Omega_{m0}, \omega_{0}, \omega_a, \gamma_0, \gamma_a \}$ and their uncertainties. 

The best-fit values of the cosmological parameters that we obtained are listed as follows: 
\begin{itemize}
\item {\it Model 1}: First, we consider the background evolution to be the same as that of the $\Lambda$CDM model. Thus, we set $\omega_{0}=-1$ and $\omega_a=0$ hence $\omega_{DE}=-1$. In addition, for simplicity, we choose $\gamma_a=0$ and find the best-fit values for $\{\Omega_{m0}, \gamma_0 \}$ = $\{0.2771, 0.5841\}$.
\item {\it Model 2}: As the second model, we investigate the $\omega$CDM model where $\omega_{DE}=\omega_0+\omega_a(1-a)$. However, we first consider the case where the EoS is constant hence $\omega_a=0$. In this case, the best-fit values we obtain are: $\{\Omega_{m0}, \omega_{0}, \gamma_0, \gamma_a \}$ = $\{0.2768, -0.9986, 0.5454, -0.0099\}$.
\end{itemize}
The third model we discuss in this section is not from the likelihood analysis. 
\begin{itemize}
\item {\it Model 3}: We adopt $\left\{ \Omega_{\m0}, \omega_{0}, \omega_{a}, \gamma_{0}, \gamma_{a} \right\} = \left\{ 0.32, -1, 0, 0.55, 0 \right\}$ to be similar to Planck data~\cite{Aghanim:2018eyx}. 
\end{itemize}

By using these observationally favored values of the cosmological parameters, we reconstruct both numerical and analytic viable $f(\tmg_{\mA})$ gravity models in the following subsection.

\subsection{Reconstructed Models}\label{subsec:models}
Based on different cosmological parameters obtained in the last subsection \ref{subsec:observation}, we reconstruct and investigate various models of $R+f(\mg)$ gravity in this subsection. As we will shortly see, the following analytic functions well describe our numerical models 
\begin{align}
\tf_1 \left( \mg_{\mA} \right) &=  \left( \tmg_{\mA} \right)^{m_1} \left[ a_1 + b_1 \left( \tmg_{\mA} \right)^{n_1} \right] \,, \label{AnalMod1} \\
\tf_2 \left( \mg_{\mA} \right) &=  \frac{ a_{2} + b_{2} \left( \tmg_{\mA} \right)^{m_2}}{ c_{2} + d_{2} \left( \tmg_{\mA} \right)^{n_2}  }\,, \label{AnalMod2} 
\end{align}
where coefficients $a_1$, $b_1$, $m_1$, $n_1$, $a_2$, $b_2$, $c_2$, $d_2$, $m_2$, and $n_2$ are obtained from different models.

\subsubsection{Model 1 : $\Lambda$CDM model} \label{subsec:LCDM}

\begin{figure}[h!]
	\centering
	\vspace{1cm}
	\begin{tabular}{cc}
	\epsfig{file=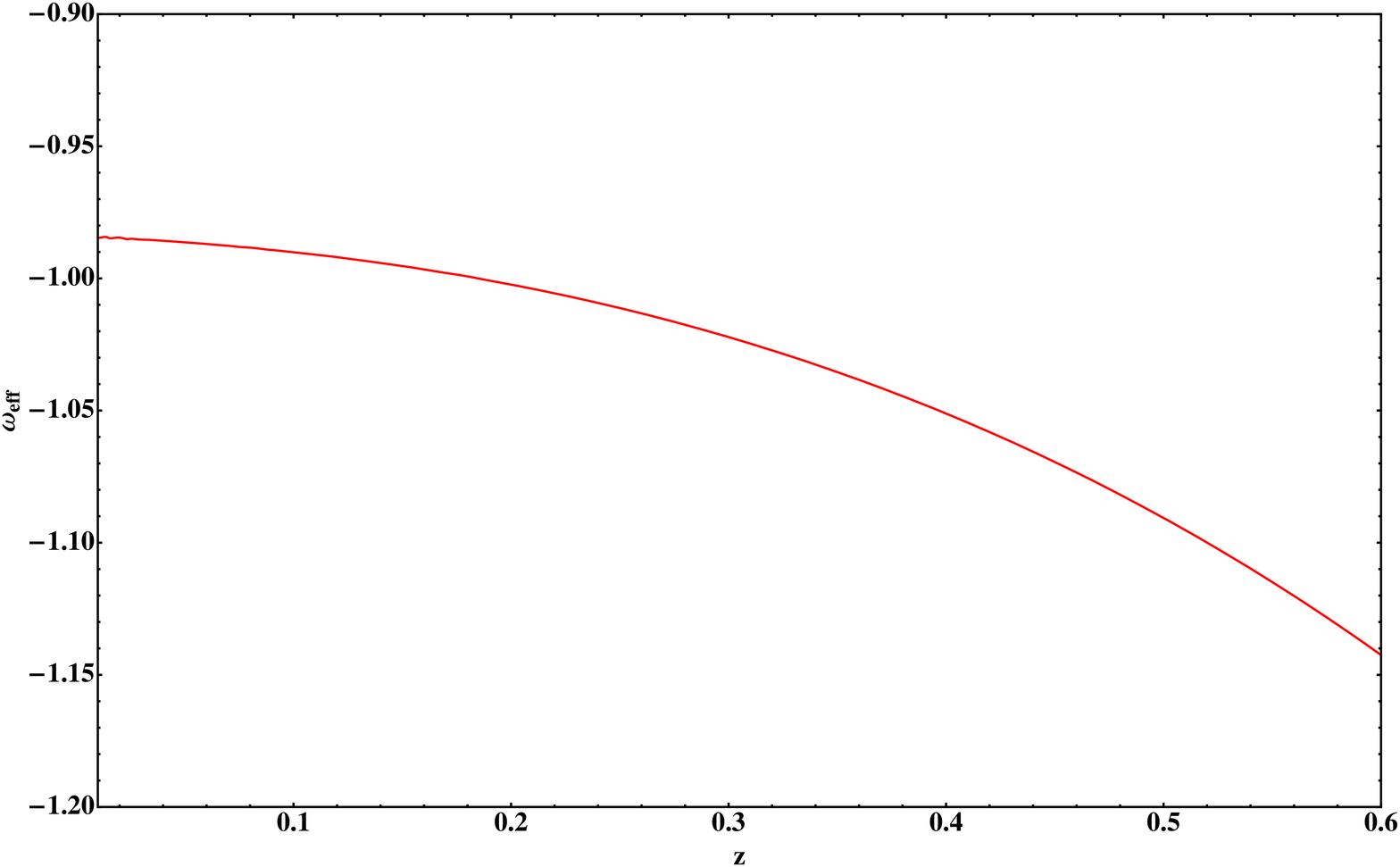,width=0.50\linewidth,clip=} &
	\epsfig{file=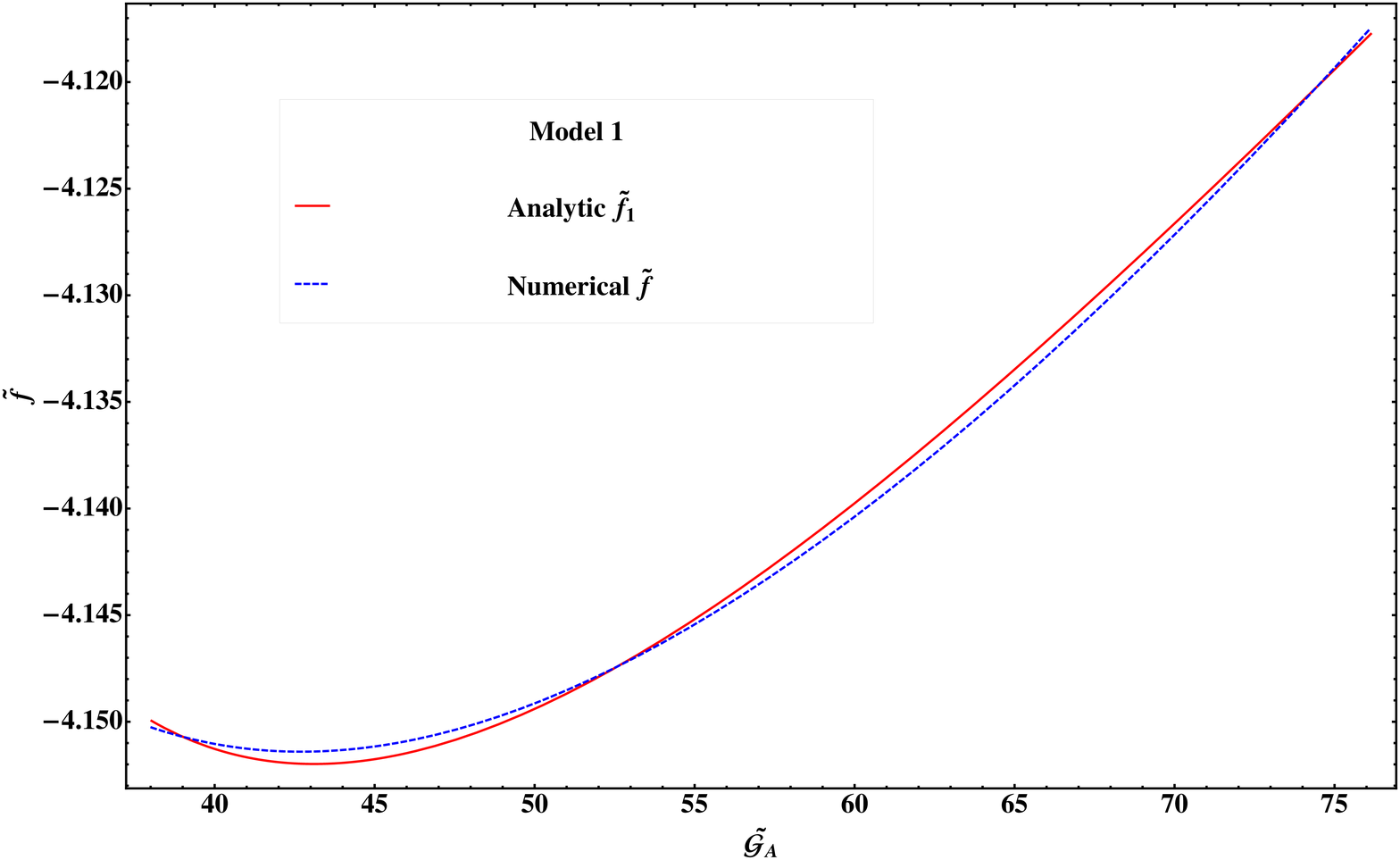,width=0.50\linewidth,clip=}
	\end{tabular}
\vspace{-0.5cm}
\caption{For {\it Model 1}: a) The red-shift evolution of $\omega_{\eff}$. b) Evolutions of the numerical and the analytic solutions as a function $\tmg_{\mA}$. The solid line indicates the analytic solution and the dashed one denotes the numerical one, respectively. } \label{Fig1}
\vspace{1cm}
\end{figure}

As we mentioned earlier, the background evolution of this model is the same as that of the $\Lambda$CDM model: {\it i.e.,} $\omega_{0} = -1$ and $\omega_{a}=0$. In addition, by choosing $\gamma_{a} = 0$, we obtained the best-fit values for $\left\{ \Omega_{\m0}, \gamma_{0} \right\} = \left\{ 0.2771, 0.5841 \right\}$ as shown in the previous subsection~\ref{subsec:observation}. In order to obtain viable numerical solutions, one needs to specify $\mA$ and $\tf_{0}^{''}$. We find that $\{\mA , \tf_{0}^{''}\} = \{1.0, 0.26\}$ produce the viable numerical solution of $\tf$. However, the stability condition seems to be challenged for $z > 0.6$ in this model because of the high values of $\gamma_0$. We find that it is difficult to obtain the stable slowly varying $\tf$ for $\gamma_0 > 0.6$. Thus, it is difficult to find viable models when we obtain cosmological parameters with large $\gamma_0$ values. The behavior of the EDE equation of state, $\omega_{\eff}$ is shown in the left panel of Fig.~\ref{Fig1}. Moreover, one can find the approximate analytic solution of this model. We find that the first analytic function, $\tf_{1}$ given in Eq.~\eqref{AnalMod1} with $\{a_1, b_1, m_1, n_1\} = \{-3.6, 2.7 \times 10^{-2}, 5.9 \times 10^{-2}, 0.65\}$ approximately mimics the numerical solution as shown in the right panel of Fig.~\ref{Fig1}. The solid line indicates the analytic solution and the dashed one denotes the numerical one, respectively. In this model, we are not able to find the second viable analytic solution, $\tf_2$ given in Eq.~\eqref{AnalMod2}. 

\subsubsection{Model 2 : $\omega$CDM model with $\omega_{0} \neq -1$ and $\omega_a=0$} \label{subsec:wCDM}

\begin{figure}[h!]
	\centering
	\vspace{1cm}
	\begin{tabular}{cc}
	\epsfig{file=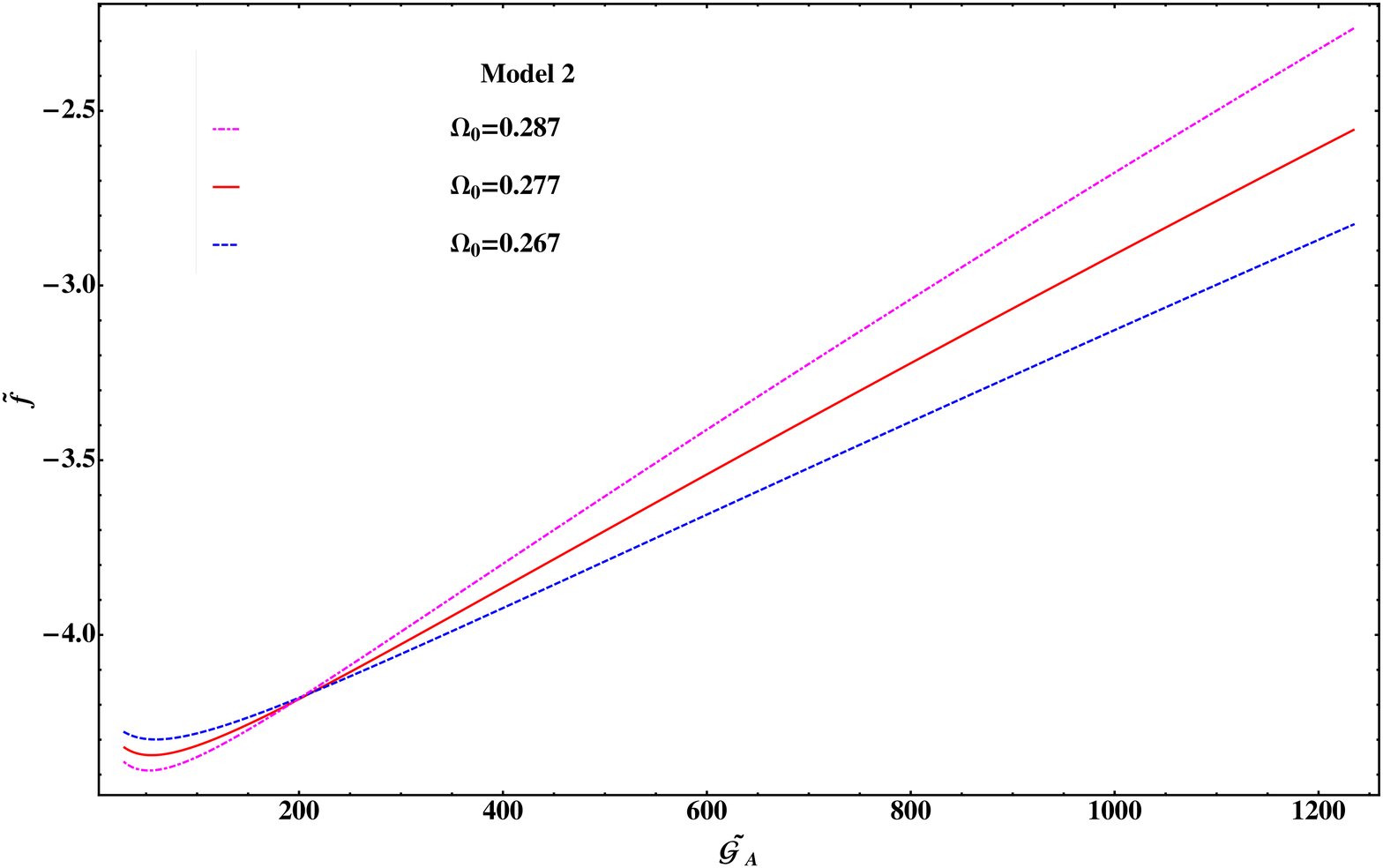,width=0.50\linewidth,clip=} &
	\epsfig{file=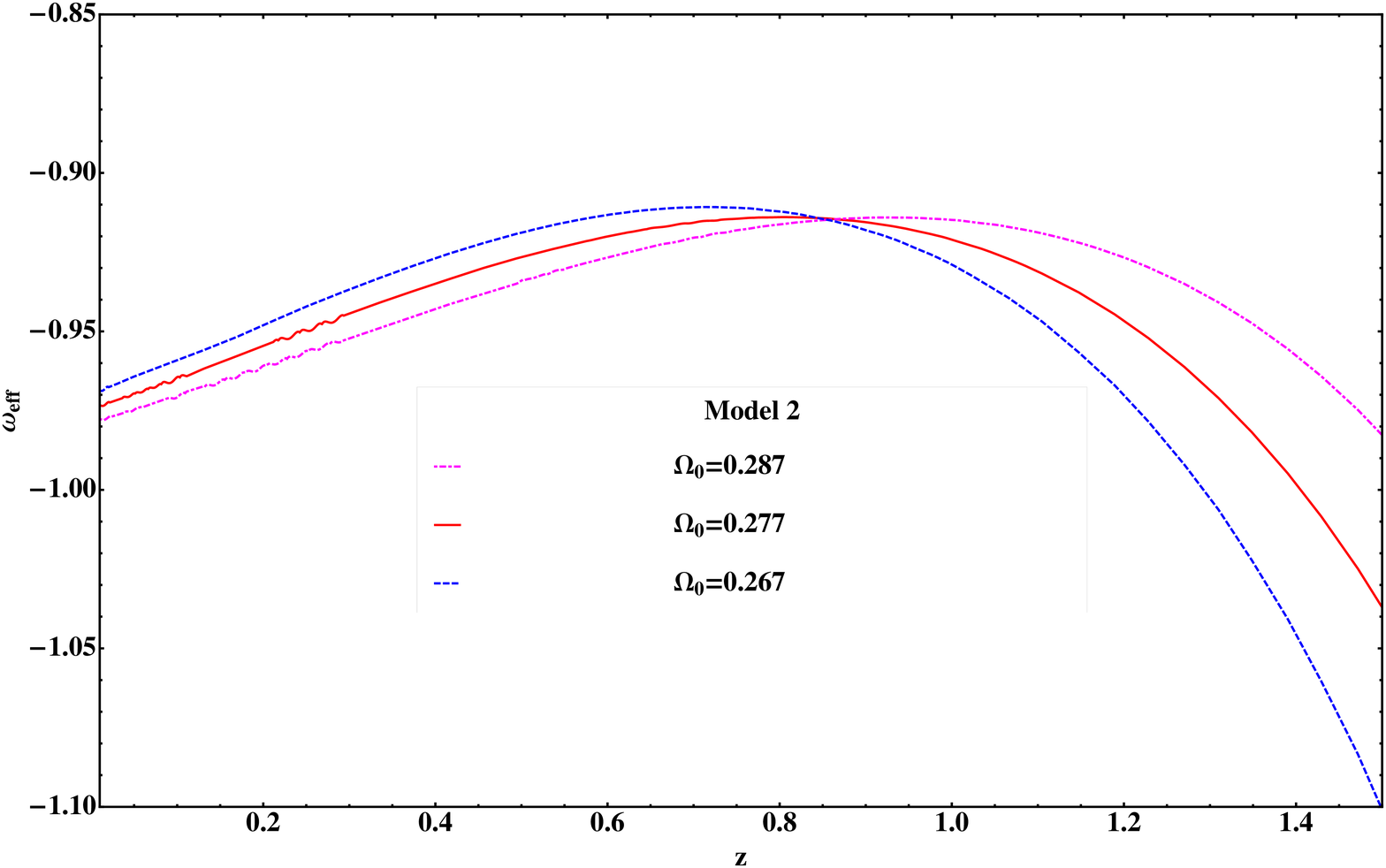,width=0.50\linewidth,clip=}
	\end{tabular}
\vspace{-0.5cm}
\caption{For {\it Model 2}: a) Evolution of different models. The dot-dashed, solid, and dashed lines correspond $\Omega_{\m0} = 0.287, 0.277$, and 0.267, respectively. b) Evolutions of $\omega_{\eff}$s for the different values of $\Omega_{\m0}$.} \label{Fig2}
\vspace{1cm}
\end{figure}

As the second model, we investigate the $\omega$CDM models with $\omega_{0} \neq -1$ and $\omega_a=0$ while allowing the time evolution of the growth index rate (${\it i.e.} \, \gamma_{a} \neq 0$) to obtain the best fit values as $\left\{ \Omega_{\m0}, \omega_{0}, \gamma_{0}, \gamma_{a} \right\} = \left\{ 0.277, -0.999,  0.545, -0.01 \right\}$. The background evolution for this model is still effectively the same as that of the $\Lambda$CDM model. However, one can obtain the smaller value of $\gamma_0$ by relaxing the condition on $\gamma_{a} \neq 0$ compared to the {\it Model 1}. This provides the viable $\tf$ model which is consistent with observations.  For the given cosmological parameters, we find that $\{\mA , \tf_{0}^{''}\} = \{0.6, -0.08\}$ produce the viable numerical solution of $\tf$. In this model, we also investigate the effects of the change in the $\Omega_{\m0}$ value by comparing cosmological evolution of $\tf$ model as a function of $\tmg_{\mA}$ for the different values of $\Omega_{\m0}$.  These are shown in the left panel of Fig.~\ref{Fig2}. 

If one increases the value of $\Omega_{\m0}$ compared to the best fit value, the slope of $\tf$ gets the larger compared to that of the best-fit value. As the value of $\Omega_{\m0}$ decreases, so does the variation of the function $\tf$. The dot-dashed, solid, and dashed lines correspond $\Omega_{\m0} = 0.287, 0.277$, and 0.267, respectively. 
We further investigate behavior of $\omega_{\eff}$ for different values of $\Omega_{\m0}$. One might expect steeper variation of $\omega_{\eff}$ for the larger value of $\Omega_{\m0}$. However, it is opposite to the expectation as shown in the right panel of Fig.~\ref{Fig2}. This is due to the fact that $\omega_{\eff}$ does not simply depend on the differentiation of $\tf$ as given in Eq.~\eqref{omegaeff}. The red-shift evolution of $\omega_{\eff}$ are depicted as dot-dashed, solid, and dashed lines for $\Omega_{\m0} = 0.287, 0.277$, and 0.267, respectively.

\begin{figure}[h!]
	\centering
	\vspace{1cm}
	\epsfig{file=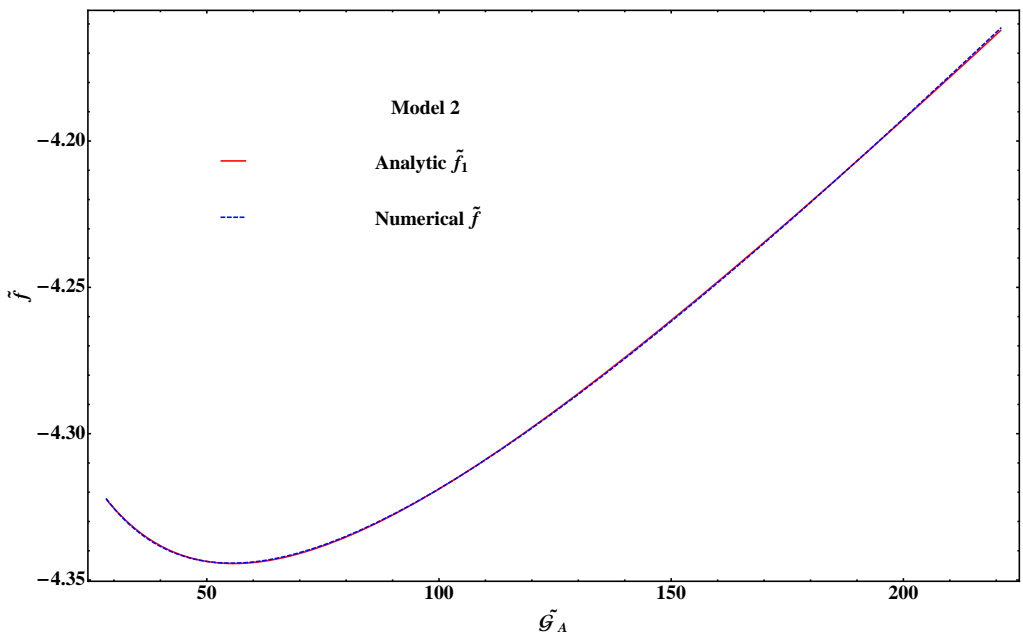,width=0.8\linewidth,clip=} 
\vspace{-0.5cm}
\caption{The comparison of analytic forms $\tf_1$ with the numerical solution $\tf$ for {\it Model 2}. } \label{Fig3}
\vspace{1cm}
\end{figure}

Also for this model, we can find the analytic solution which is well matched with the numerical one. We find that the first analytic function, $\tf_{1}$ given in Eq.~\eqref{AnalMod1} with $\{a_1, b_1, m_1, n_1\} = \{-4.0, 5.2 \times 10^{-3}, 3.3 \times 10^{-2}, 0.84\}$ almost perfectly matches with the numerical solution shown in Fig.~\ref{Fig3}. The solid and dashed lines correspond analytic and numerical solutions, respectively. In this model, we could not find the viable solution that described by the second analytic function, $\tf_2$ given in Eq.~\eqref{AnalMod2}.    

\subsubsection{Model 3: $\Lambda$CDM with $\Omega_{\m0} =0.32$ } \label{subsec:Planck}
\begin{figure}[h!]
	\centering
	\vspace{1cm}
	\begin{tabular}{cc}
	\epsfig{file=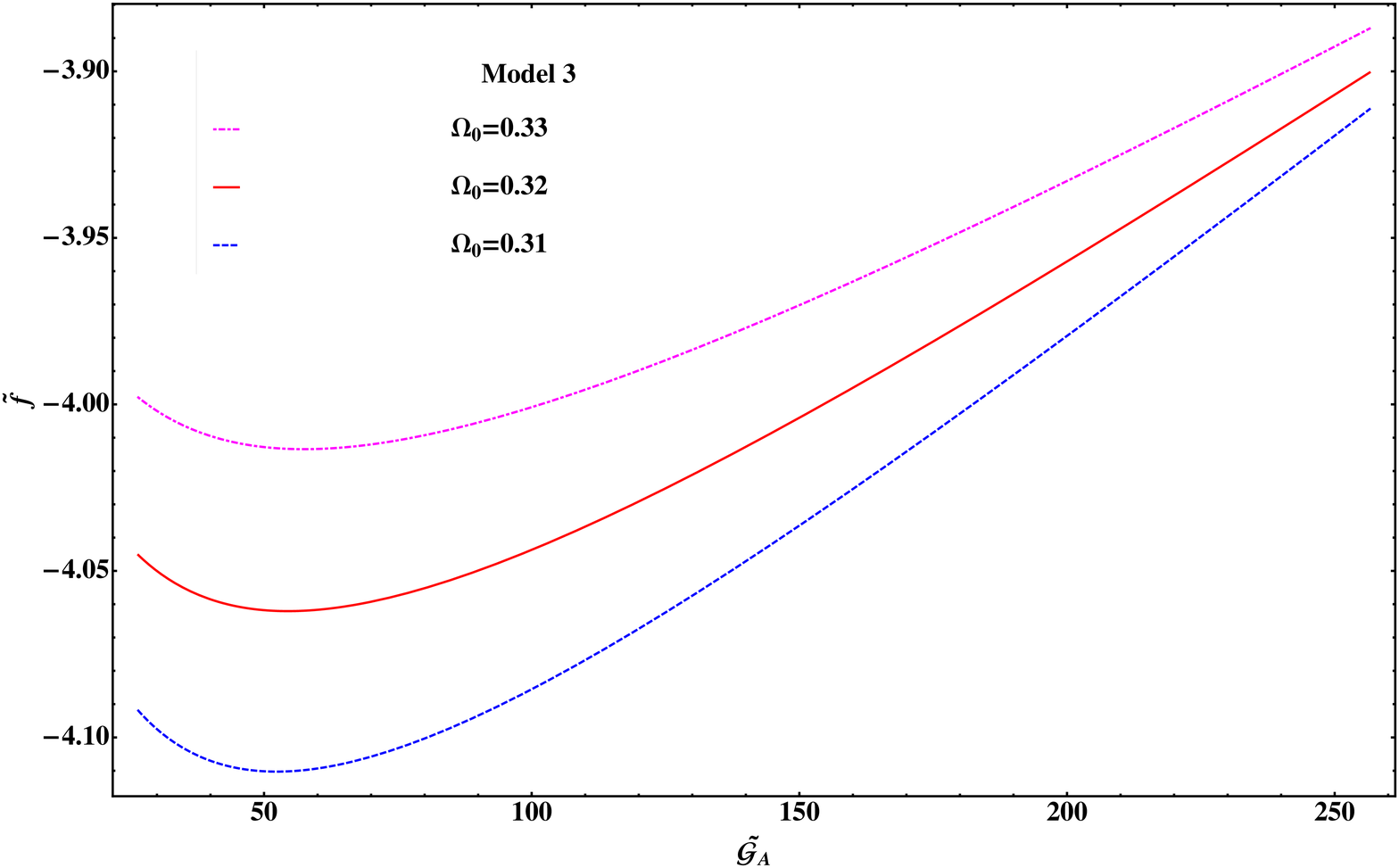,width=0.50\linewidth,clip=} &
	\epsfig{file=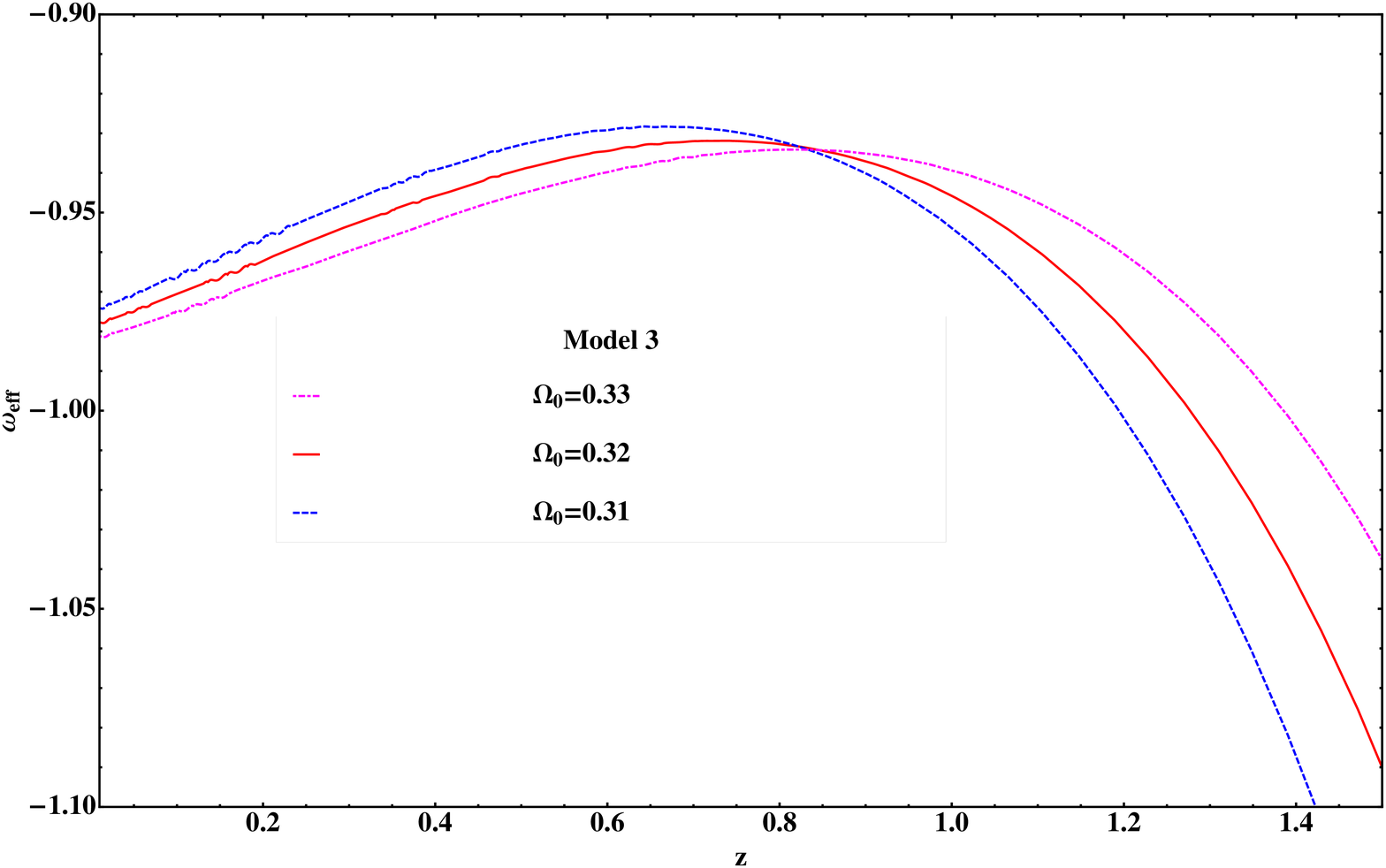,width=0.50\linewidth,clip=}
	\end{tabular}
\vspace{-0.5cm}
\caption{For {\it Model 3}: a) Evolution of different models as a function of $\tmg_{\mA}$. The dot-dashed, solid, and dashed lines correspond $\Omega_{\m0} = 0.33, 0.32$, and 0.31, respectively.  b) Evolution of $\omega_{\eff}$ for different value of $\Omega_{\m0}$.} \label{Fig4}
\vspace{1cm}
\end{figure}

We investigate the model with values of cosmological parameters similar to that of Planck~\cite{Aghanim:2018eyx}. For this purpose, we adopt $\left\{ \Omega_{\m0}, \omega_{0}, \omega_{a}, \gamma_{0}, \gamma_{a} \right) = \left( 0.32, -1, 0, 0.55, 0 \right\}$. For the given cosmological parameters, we find that one can obtain viable numerical solutions for $\{\mA , \tf_{0}^{''}\}= \{0.6, -0.05\}$. With these initial conditions, one can reconstruct the cosmological evolution of $\tf$ model numerically by using Eq.~\eqref{MEdimf}. By comparing cosmological evolution of $\tf$ model as a function of $\tmg_{\mA}$ with varying the $\Omega_{\m0}$ values from 0.31 to 0.33, we also investigate the effects of different values of $\Omega_{\m0}$. These are shown in the left panel of Fig.~\ref{Fig4}. The larger the $\Omega_{\m0}$ values, the smaller the change in $\tf$ for this model. The dot-dashed, solid, and dashed lines correspond $\Omega_{\m0} = 0.33, 0.32$, and 0.31, respectively. The red-shift evolution of $\omega_{\eff}$ for different values of $\Omega_{\m0}$ is also investigated. This is shown in the right panel of Fig.~\ref{Fig4}. The red-shift evolution of $\omega_{\eff}$ are depicted as dot-dashed, solid, and dashed lines for $\Omega_{\m0} = 0.33, 0.32$, and 0.31, respectively. 

\begin{figure}[h!]
	\centering
	\vspace{1cm}
	\begin{tabular}{cc}
	\epsfig{file=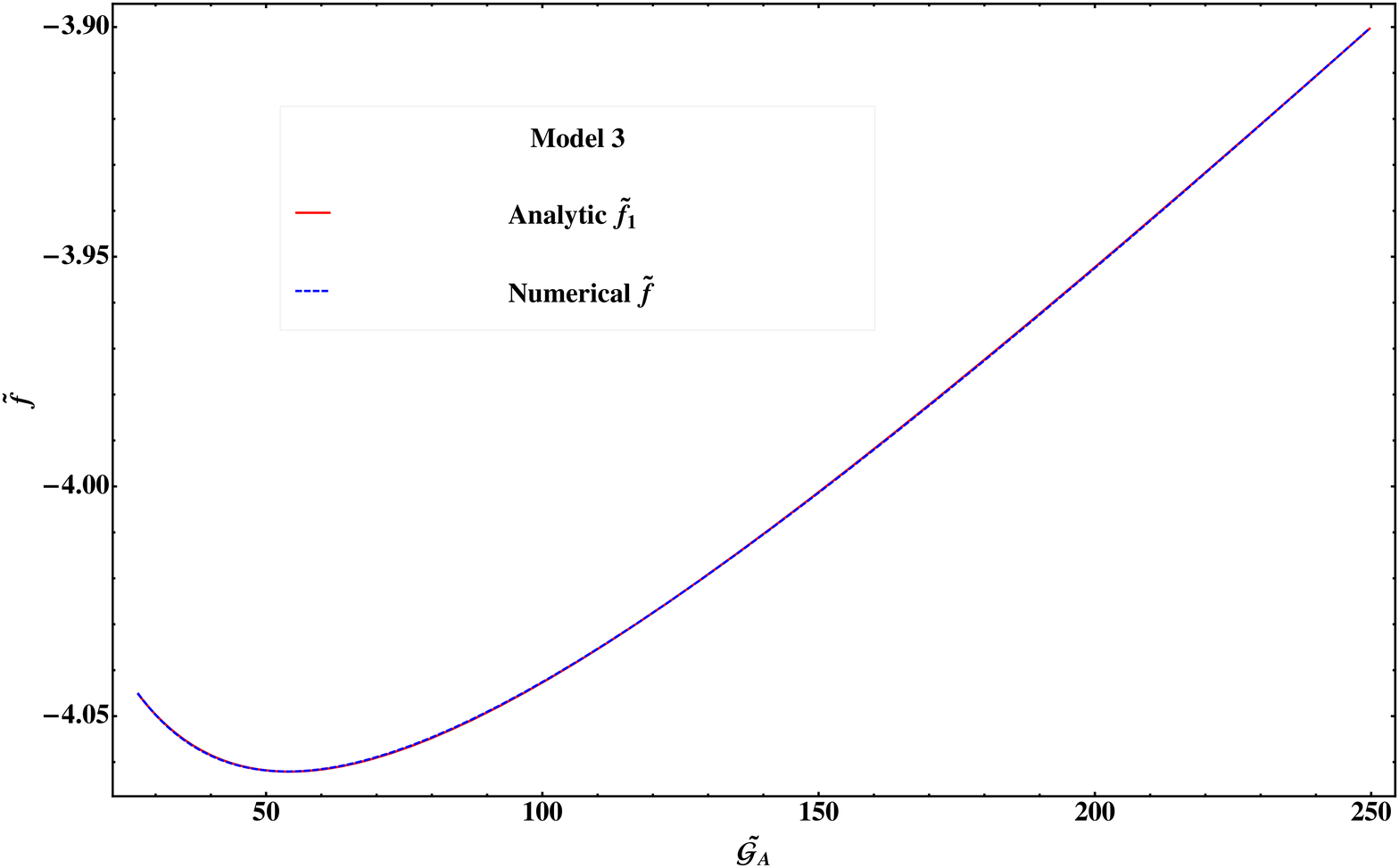,width=0.50\linewidth,clip=} &
	\epsfig{file=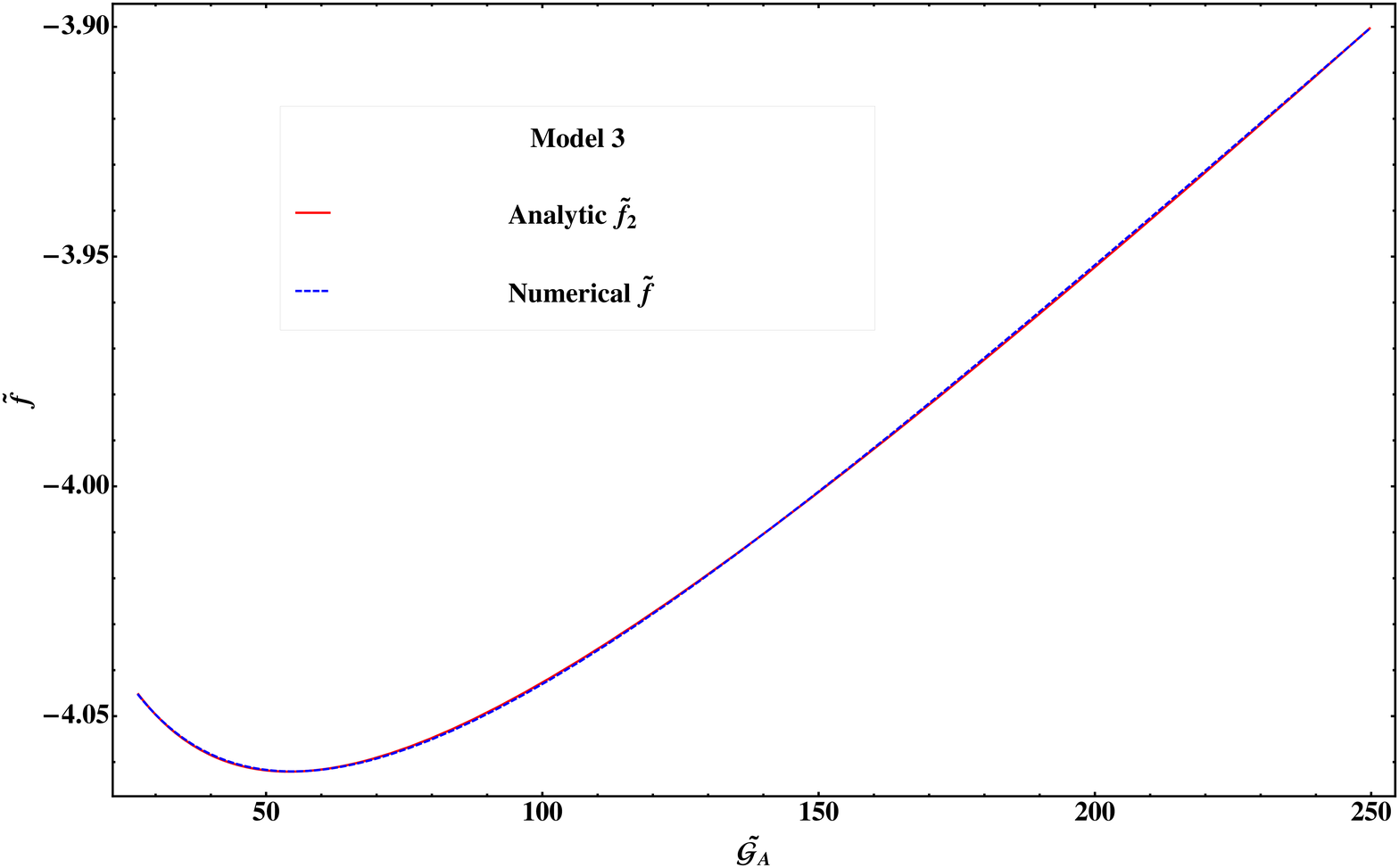,width=0.50\linewidth,clip=}
	\end{tabular}
\vspace{-0.5cm}
\caption{The comparison of analytic forms $\tf_1$ and $\tf_2$ with the numerical solutions of $\tf$. a) The solid and dashed lines correspond $\tf_1$ and $\tf$, respectively. b) The solid and dashed lines correspond $\tf_2$ and $\tf$, respectively. } \label{Fig5}
\vspace{1cm}
\end{figure}

For this model, one can find the analytic solutions which are almost identical to the numerical ones. We find that the first analytic function, $\tf_{1}$ given in Eq.~\eqref{AnalMod1} with $\{a_1, b_1, m_1, n_1\} = \{-4.0, 5.2 \times 10^{-3}, 3.3 \times 10^{-2}, 0.84\}$ almost perfectly matches with the numerical solution shown in the left panel of Fig.~\ref{Fig5}. The solid and dashed lines correspond analytic and numerical solutions, respectively. We could also obtained the second viable analytic solution, $\tf_2$ given in Eq.~\eqref{AnalMod2} with $\{a_2, b_2, c_2, d_2, m_2, n_2\} = \{-314, -8.8, 82.9, 0.23, 0.41, 0.81\}$. This is shown in the right panel of Fig.~\ref{Fig5}. The solid and dashed lines correspond to analytic and numerical solutions, respectively. These two analytic solutions well describe the numerical ones.

\section{Conclusions}\label{sec:conclusion}
In this work, we have investigated the viable cosmological models of $f(\mg)$ gravity via a reconstruction method and presented the analytic solutions that well describe our results. After providing a brief review of both background and perturbation equations and the stability condition of the model, we have rearranged necessary equations in terms of the dimensionless quantities in Sec.~\ref{sec:dimensionless}. Thus, our setup is well prepared for the numerical investigation. However, unlike $f(R)$ gravity models, $f(\mg)$ gravity models face an unphysical challenge which leads to an occurrence of the divergence in the numerical study.  In order to cure such troublesome behavior in Sec.~\ref{sec:reconstruction}, we have generalized the original $f(\mg)$ models into the $f(\mg_\mA)$ models as given in Eq.~\eqref{mganal}, where $\mA$ is an arbitrary constant whose value is constrained by the observational data. It seems to be evident that such an extension would not affect the essence of the background as well as the perturbation evolution as it can be regarded as the simple redefinition of $\mg$, the Gauss-Bonnet term.

As an alternative to the cosmological constant in the $\Lambda$CDM model of the universe, we regard the $f(\mg_\mA)$ gravity models as dark energy with an effective equation of state $\omega_\text{eff}$, which is given in Eq.~\eqref{omegaeff}. In order to connect the model with the observations in Sec.~\ref{sec:connection}, we have used the so-called CPL parameterization for the equation-of-state parameter and the similar form for the growth-rate index in our analysis. By employing the several observational data including CMB, Supernovae type Ia (SnIa), BAO, Hubble expansion $H(z)$, and the growth-rate data~\cite{Ade:2015xua, Wang:2015tua, Scolnic:2017caz, Sagredo:2018ahx, Beutler:2011hx, Xu:2012hg, Moresco:2016mzx, Guo:2015gpa, Anderson:2013zyy, Ross:2014qpa, Gil-Marin:2015nqa, Amendola:2012wc} together with the statistical methods based on $\chi_{tot}^{2}$, we have obtained the best-fit values of the cosmological parameters in Sec.~\ref{subsec:observation}. These best-fit values have been used in Sec.~\ref{subsec:models} for reconstructing the viable $f(\mg_\mA)$ gravity models. As a result of our numerical analysis, we have successfully reconstructed the cosmological models of $f(\mg_\mA)$ gravity that well describe the observational data. Moreover, in Eqs.~\eqref{AnalMod1} and ~\eqref{AnalMod2}, we have provided the analytic functions that almost perfectly match with our numerical results by using the different set of best-fit and observationally favored values,  see Figs.~\ref{Fig1} -- \ref{Fig5} and their interpretations in the main text. It is therefore worth investigating the physical origin of those solutions and their cosmological implications, which we leave as future extensions to our present study.

\section*{Acknowledgments}
GT would like to appreciate Sungkyunkwan University for its hospitality during the completion of this work. SL is supported by Basic Science Research Program through the National Research Foundation of Korea (NRF) funded by the Ministry of Science, ICT and Future Planning (Grant No. NRF-2017R1A2B4011168). GT was supported by IBS under the project code, IBS-R018-D1. 




\begin{thebibliography}{99}

\bibitem{Riess:1998cb} 
  A.~G.~Riess {\it et al.} [Supernova Search Team],
  ``Observational evidence from supernovae for an accelerating universe and a cosmological constant,''
  Astron.\ J.\  {\bf 116}, 1009 (1998);
  
\bibitem{Perlmutter:1998np} 
  S.~Perlmutter {\it et al.} [Supernova Cosmology Project Collaboration],
  ``Measurements of $\Omega$ and $\Lambda$ from 42 high redshift supernovae,''
  Astrophys.\ J.\  {\bf 517}, 565 (1999);
  
\bibitem{Spergel:2003cb} 
  D.~N.~Spergel {\it et al.} [WMAP Collaboration],
  ``First year Wilkinson Microwave Anisotropy Probe (WMAP) observations: Determination of cosmological parameters,''
  Astrophys.\ J.\ Suppl.\  {\bf 148}, 175 (2003);
  
\bibitem{Hinshaw:2012aka} 
  G.~Hinshaw {\it et al.} [WMAP Collaboration],
  ``Nine-Year Wilkinson Microwave Anisotropy Probe (WMAP) Observations: Cosmological Parameter Results,''
  Astrophys.\ J.\ Suppl.\  {\bf 208}, 19 (2013);
 
\bibitem{Ade:2013zuv} 
  P.~A.~R.~Ade {\it et al.} [Planck Collaboration],
  ``Planck 2013 results. XVI. Cosmological parameters,''
  Astron.\ Astrophys.\  {\bf 571}, A16 (2014);
  
\bibitem{Ade:2015xua} 
  P.~A.~R.~Ade {\it et al.} [Planck Collaboration],
  ``Planck 2015 results. XIII. Cosmological parameters,''
  Astron.\ Astrophys.\  {\bf 594}, A13 (2016);
  
\bibitem{Aghanim:2018eyx} 
  N.~Aghanim {\it et al.} [Planck Collaboration],
  ``Planck 2018 results. VI. Cosmological parameters,''
  arXiv:1807.06209 [astro-ph.CO];
  
\bibitem{Eisenstein:2005su} 
  D.~J.~Eisenstein {\it et al.} [SDSS Collaboration],
  ``Detection of the Baryon Acoustic Peak in the Large-Scale Correlation Function of SDSS Luminous Red Galaxies,''
  Astrophys.\ J.\  {\bf 633}, 560 (2005);

\bibitem{Sahni:1999gb} 
  V.~Sahni and A.~A.~Starobinsky,
  ``The Case for a positive cosmological Lambda term,''
  Int.\ J.\ Mod.\ Phys.\ D {\bf 9}, 373 (2000);
  
  P.~J.~E.~Peebles and B.~Ratra,
  ``The Cosmological constant and dark energy,''
  Rev.\ Mod.\ Phys.\  {\bf 75}, 559 (2003);

\bibitem{Carroll:2000fy} 
  S.~M.~Carroll,
  ``The Cosmological constant,''
  Living Rev.\ Rel.\  {\bf 4}, 1 (2001);
  
\bibitem{DeFelice:2010aj} 
  A.~De Felice and S.~Tsujikawa,
  ``f(R) theories,''
  Living Rev.\ Rel.\  {\bf 13}, 3 (2010);
  
\bibitem{Linder:2010py} 
  E.~V.~Linder,
  ``Einstein's Other Gravity and the Acceleration of the Universe,''
  Phys.\ Rev.\ D {\bf 81}, 127301 (2010);
  Erratum: [Phys.\ Rev.\ D {\bf 82}, 109902 (2010)];
  
\bibitem{Clifton:2011jh} 
  T.~Clifton, P.~G.~Ferreira, A.~Padilla and C.~Skordis,
  ``Modified Gravity and Cosmology,''
  Phys.\ Rept.\  {\bf 513}, 1 (2012);
  
\bibitem{Mortonson:2013zfa} 
  M.~J.~Mortonson, D.~H.~Weinberg and M.~White,
  ``Dark Energy: A Short Review,''
  arXiv:1401.0046 [astro-ph.CO];
  
\bibitem{Koyama:2015vza} 
  K.~Koyama,
  ``Cosmological Tests of Modified Gravity,''
  Rept.\ Prog.\ Phys.\  {\bf 79}, no. 4, 046902 (2016);
  
\bibitem{Capozziello:2002rd} 
  S.~Capozziello,
  ``Curvature quintessence,''
  Int.\ J.\ Mod.\ Phys.\ D {\bf 11}, 483 (2002);
  
\bibitem{Carroll:2003wy} 
  S.~M.~Carroll, V.~Duvvuri, M.~Trodden and M.~S.~Turner,
  ``Is cosmic speed - up due to new gravitational physics?,''
  Phys.\ Rev.\ D {\bf 70}, 043528 (2004);
 
\bibitem{Nojiri:2003ni} 
  S.~Nojiri and S.~D.~Odintsov,
  ``Modified gravity with ln R terms and cosmic acceleration,''
  Gen.\ Rel.\ Grav.\  {\bf 36}, 1765 (2004);

\bibitem{Easson:2005ax} 
  D.~A.~Easson, F.~P.~Schuller, M.~Trodden and M.~N.~R.~Wohlfarth,
  ``Cosmological constraints on a classical limit of quantum gravity,''
  Phys.\ Rev.\ D {\bf 72}, 043504 (2005);
 
 
\bibitem{Carroll:2004de} 
  S.~M.~Carroll, A.~De Felice, V.~Duvvuri, D.~A.~Easson, M.~Trodden and M.~S.~Turner,
  ``The Cosmology of generalized modified gravity models,''
  Phys.\ Rev.\ D {\bf 71}, 063513 (2005);
  
\bibitem{Allemandi:2004wn} 
  G.~Allemandi, A.~Borowiec and M.~Francaviglia,
  ``Accelerated cosmological models in Ricci squared gravity,''
  Phys.\ Rev.\ D {\bf 70}, 103503 (2004);
 
  
\bibitem{Nojiri:2005jg} 
  S.~Nojiri and S.~D.~Odintsov,
  ``Modified Gauss-Bonnet theory as gravitational alternative for dark energy,''
  Phys.\ Lett.\ B {\bf 631}, 1 (2005);

\bibitem{Cognola:2006eg} 
  G.~Cognola, E.~Elizalde, S.~Nojiri, S.~D.~Odintsov and S.~Zerbini,
  ``Dark energy in modified Gauss-Bonnet gravity: Late-time acceleration and the hierarchy problem,''
  Phys.\ Rev.\ D {\bf 73}, 084007 (2006);
  
\bibitem{Bamba:2009uf} 
  K.~Bamba, S.~D.~Odintsov, L.~Sebastiani and S.~Zerbini,
  ``Finite-time future singularities in modified Gauss-Bonnet and F(R,G) gravity and singularity avoidance,''
  Eur.\ Phys.\ J.\ C {\bf 67}, 295 (2010);
 
 
\bibitem{Li:2007jm} 
  B.~Li, J.~D.~Barrow and D.~F.~Mota,
  ``The Cosmology of Modified Gauss-Bonnet Gravity,''
  Phys.\ Rev.\ D {\bf 76}, 044027 (2007);

\bibitem{Cognola:2006sp} 
  G.~Cognola, E.~Elizalde, S.~Nojiri, S.~Odintsov and S.~Zerbini,
  ``String-inspired Gauss-Bonnet gravity reconstructed from the universe expansion history and yielding the transition from matter dominance to dark energy,''
  Phys.\ Rev.\ D {\bf 75}, 086002 (2007);
 
\bibitem{Nojiri:2007bt} 
  S.~Nojiri, S.~D.~Odintsov and P.~V.~Tretyakov,
  ``From inflation to dark energy in the non-minimal modified gravity,''
  Prog.\ Theor.\ Phys.\ Suppl.\  {\bf 172}, 81 (2008);

\bibitem{DeFelice:2008wz} 
  A.~De Felice and S.~Tsujikawa,
  ``Construction of cosmologically viable f(G) dark energy models,''
  Phys.\ Lett.\ B {\bf 675}, 1 (2009);
 
 
\bibitem{Zhou:2009cy} 
  S.~Y.~Zhou, E.~J.~Copeland and P.~M.~Saffin,
  ``Cosmological Constraints on $f(G)$ Dark Energy Models,''
  JCAP {\bf 0907}, 009 (2009);
 
\bibitem{Uddin:2009wp}
  K.~Uddin, J.~E.~Lidsey and R.~Tavakol,
  ``Cosmological scaling solutions in generalised Gauss-Bonnet gravity theories,''
  Gen.\ Rel.\ Grav.\  {\bf 41} (2009) 2725;
 
\bibitem{DeFelice:2009aj} 
  A.~De Felice and S.~Tsujikawa,
  ``Solar system constraints on f(G) gravity models,''
  Phys.\ Rev.\ D {\bf 80}, 063516 (2009);
   
\bibitem{Linder:2004ng} 
  E.~V.~Linder,
  ``Probing gravitation, dark energy, and acceleration,''
  Phys.\ Rev.\ D {\bf 70}, 023511 (2004);
 
\bibitem{Linder:2007hg} 
  E.~V.~Linder and R.~N.~Cahn,
  ``Parameterized Beyond-Einstein Growth,''
  Astropart.\ Phys.\  {\bf 28}, 481 (2007);
 
\bibitem{Steigerwald:2014ava} 
  H.~Steigerwald, J.~Bel and C.~Marinoni,
  ``Probing non-standard gravity with the growth index: a background independent analysis,''
  JCAP {\bf 1405}, 042 (2014);
 
\bibitem{Basilakos:2017rgc} 
  S.~Basilakos and S.~Nesseris,
  ``Conjoined constraints on modified gravity from the expansion history and cosmic growth,''
  Phys.\ Rev.\ D {\bf 96}, no. 6, 063517 (2017);


\bibitem{Carloni:2010ph} 
  S.~Carloni, R.~Goswami and P.~K.~S.~Dunsby,
  ``A new approach to reconstruction methods in $f(R)$ gravity,''
  Class.\ Quant.\ Grav.\  {\bf 29}, 135012 (2012);
 
\bibitem{He:2012rf} 
  J.~h.~He and B.~Wang,
  ``Revisiting $f(R)$ gravity models that reproduce $\Lambda$CDM expansion,''
  Phys.\ Rev.\ D {\bf 87}, no. 2, 023508 (2013);
 
\bibitem{Xu:2014wda} 
  L.~Xu,
  ``Constraint on $f(R)$ Gravity through the Redshift Space Distortion,''
  Phys.\ Rev.\ D {\bf 91}, no. 6, 063008 (2015);
  
\bibitem{Lee:2017lud} 
  S.~Lee,
  ``Reconstruction of f(R) gravity models from observations,''
  Phys.\ Dark Univ.\  {\bf 25}, 100305 (2019);
  
\bibitem{Chevallier:2000qy} 
  M.~Chevallier and D.~Polarski,
  ``Accelerating universes with scaling dark matter,''
  Int.\ J.\ Mod.\ Phys.\ D {\bf 10}, 213 (2001);
  
\bibitem{Linder:2002et} 
  E.~V.~Linder,
  ``Exploring the expansion history of the universe,''
  Phys.\ Rev.\ Lett.\  {\bf 90}, 091301 (2003);
  
  
\bibitem{DeFelice:2010hb} 
  A.~De Felice and T.~Suyama,
  ``Linear growth of matter density perturbations in f(R,G) theories,''
  Prog.\ Theor.\ Phys.\  {\bf 125}, 603 (2011);
  
\bibitem{Wang:2015tua} 
  Y.~Wang and M.~Dai,
  ``Exploring uncertainties in dark energy constraints using current observational data with Planck 2015 distance priors,''
  Phys.\ Rev.\ D {\bf 94}, no. 8, 083521 (2016);

\bibitem{Scolnic:2017caz} 
  D.~M.~Scolnic {\it et al.},
  ``The Complete Light-curve Sample of Spectroscopically Confirmed SNe Ia from Pan-STARRS1 and Cosmological Constraints from the Combined Pantheon Sample,''
  Astrophys.\ J.\  {\bf 859}, no. 2, 101 (2018);

\bibitem{Beutler:2011hx} 
  F.~Beutler {\it et al.},
  ``The 6dF Galaxy Survey: Baryon Acoustic Oscillations and the Local Hubble Constant,''
  Mon.\ Not.\ Roy.\ Astron.\ Soc.\  {\bf 416}, 3017 (2011);
  
\bibitem{Xu:2012hg} 
  X.~Xu, N.~Padmanabhan, D.~J.~Eisenstein, K.~T.~Mehta and A.~J.~Cuesta,
  ``A 2
  Mon.\ Not.\ Roy.\ Astron.\ Soc.\  {\bf 427}, 2146 (2012);


\bibitem{Anderson:2013zyy} 
  L.~Anderson {\it et al.} [BOSS Collaboration],
  ``The clustering of galaxies in the SDSS-III Baryon Oscillation Spectroscopic Survey: baryon acoustic oscillations in the Data Releases 10 and 11 Galaxy samples,''
  Mon.\ Not.\ Roy.\ Astron.\ Soc.\  {\bf 441}, no. 1, 24 (2014);
  
\bibitem{Ross:2014qpa} 
  A.~J.~Ross, L.~Samushia, C.~Howlett, W.~J.~Percival, A.~Burden and M.~Manera,
  ``The clustering of the SDSS DR7 main Galaxy sample – I. A 4 per cent distance measure at $z = 0.15$,''
  Mon.\ Not.\ Roy.\ Astron.\ Soc.\  {\bf 449}, no. 1, 835 (2015);
  
\bibitem{Gil-Marin:2015nqa} 
  H.~Gil-Marín {\it et al.},
  ``The clustering of galaxies in the SDSS-III Baryon Oscillation Spectroscopic Survey: BAO measurement from the LOS-dependent power spectrum of DR12 BOSS galaxies,''
  Mon.\ Not.\ Roy.\ Astron.\ Soc.\  {\bf 460}, no. 4, 4210 (2016);

\bibitem{Moresco:2016mzx} 
  M.~Moresco {\it et al.},
  ``A 6
  JCAP {\bf 1605}, 014 (2016);
  
  
\bibitem{Guo:2015gpa} 
  R.~Y.~Guo and X.~Zhang,
  ``Constraining dark energy with Hubble parameter measurements: an analysis including future redshift-drift observations,''
  Eur.\ Phys.\ J.\ C {\bf 76}, no. 3, 163 (2016);
  
  
\bibitem{Sagredo:2018ahx} 
  B.~Sagredo, S.~Nesseris and D.~Sapone,
  ``Internal Robustness of Growth Rate data,''
  Phys.\ Rev.\ D {\bf 98}, no. 8, 083543 (2018);
  
  
\bibitem{Amendola:2012wc} 
  L.~Amendola, V.~Marra and M.~Quartin,
  ``Internal Robustness: systematic search for systematic bias in SN Ia data,''
  Mon.\ Not.\ Roy.\ Astron.\ Soc.\  {\bf 430}, 1867 (2013).
  
  
\end{thebibliography}
\end{document}